\documentclass[prd,aps, preprint,  nofootinbib,superscriptaddress, 11pt]{revtex4-1}
 
\usepackage{amsmath}
\usepackage{amssymb}
\usepackage{epsfig}
\usepackage{subfigure}
\usepackage{slashed}
\usepackage{graphicx}
\usepackage{soul}
\usepackage{url}

\begin{document}

\title{Light NMSSM Neutralino Dark Matter \\  in the Wake of CDMS II and a 126 GeV Higgs}

\author{Jonathan Kozaczuk}
\email{jkozaczu@ucsc.edu}\affiliation{Department of Physics, University of California, 1156 High St., Santa Cruz, CA 95064, USA}

\author{Stefano Profumo}
\email{profumo@ucsc.edu}\affiliation{Department of Physics, University of California, 1156 High St., Santa Cruz, CA 95064, USA}\affiliation{Santa Cruz Institute for Particle Physics, Santa Cruz, CA 95064, USA} 

\date{August 28, 2013}

\begin{abstract}

\noindent Recent results from the Cryogenic Dark Matter Search (CDMS) experiment have renewed interest in light (5--15 GeV) dark matter (DM) with a large spin-independent neutralino-nucleon scattering cross-section, $\sigma_{\rm SI}\gtrsim 10^{-42}\ {\rm cm}^2$.  Previous work had suggested that the lightest neutralino in the Next-to-MSSM can fall in this mass range and achieve both the correct thermal relic abundance and the desired level for the scattering cross section, provided light Higgs bosons to mediate the pair annihilation and neutralino-nucleon scattering.  However, the requirement of a 126 GeV Standard Model-like Higgs boson significantly impacts the allowed parameter space.  Here, we examine the regions of the 
NMSSM capable of producing a light neutralino with $\sigma_{\rm SI}\sim10^{-42}$ -- 10$^{-41}$ cm$^2$, with the scattering mediated by a very light singlet-like scalar, and a 126 GeV Standard Model-like Higgs consistent with the LHC results, while satisfying other relevant cosmological, flavor and collider constraints.  We focus on two different scenarios for annihilation in the early universe, namely annihilation mediated by (1) a light scalar or by (2) a light pseudo-scalar.  As expected, both cases are highly constrained.  Nevertheless, we find that there persists potentially viable parameter space to accommodate either scenario.  In the first, accidental cancellations in the couplings allow for a SM-like Higgs with a total width and invisible branching fraction compatible with the observed Higgs boson.  Alternatively, the second scenario can occur in regions featuring smaller branching fractions of the SM-like Higgs to pairs of light scalars, pseudoscalars, and neutralinos without cancellations.  The strongest constraints in both cases come from rare meson decays and exotic decays of the SM-like Higgs boson into neutralinos and light, CP-even Higgs pairs.  We outline the relevant parameter space for both scenarios and comment on prospects for future discovery with various experiments.

\end{abstract}

\maketitle

\section{Introduction}\label{sec:intro}

The pressing question of the fundamental nature of dark matter might soon be addressed by current and near-future experimental searches. A comprehensive search strategy  that includes direct and indirect searches as well as collider probes and astronomical observations is closing in on many theoretically well-motivated dark matter particle candidates. Among the latter, WIMPs (an acronym for weakly interacting massive particles) stand out as especially compelling for a variety of reasons, not least the fact that they can naturally have a thermal relic abundance matching the observed density of dark matter in the universe.

WIMPs are predicted in numerous frameworks for physics beyond the Standard Model (SM) of elementary particles. First and foremost, many incarnations of low-scale supersymmetry (SUSY) include a WIMP candidate in the form of the lightest neutralino, which has, in fact, long been considered {\em the} prototypical WIMP. The absence thus far of any signal of supersymmetry at the Large Hadron Collider (LHC) has profound implications for the structure and origin of weak-scale SUSY, if indeed supersymmetry is realized in nature. However, LHC constraints on supersymmetric particles have so far had a relatively marginal impact on neutralinos as WIMP dark matter candidates.

Interestingly, a variety of puzzling observations have accumulated recently that all point to a WIMP candidate with a mass of about 10 GeV (for a review of such observations see Ref.~\cite{Hooper:2012ft}). These clues include the gamma-ray emission from the center of the Galaxy \cite{Hooper:2011ti}, the radio emission from certain Galactic filamentary  structures \cite{Linden:2011au}, as well as extragalactic radio emission \cite{Hooper:2012jc}, and signals reported by several direct detection experiments, including DAMA/LIBRA \cite{Bernabei:2010mq}, CoGeNT \cite{Aalseth:2011wp, Aalseth:2010vx}, CRESST-II \cite{Angloher:2011uu} and, most recently, CDMS II \cite{cdms3events}. While some of these observations might find astrophysical or instrumental explanations, and while some of these signals are in apparent conflict with other experimental results, it is certainly intriguing if not suggestive that they all point to the same mass range, in the vicinity of 10 GeV.

In particular, the 3 WIMP candidate events reported by the CDMS II experiment have attracted a good deal of attention. With an expected total background of 0.7 events, the probability of detecting three or more events is 5.4\%. However the CDMS II Collaboration reports that, taking into account the measured recoil energies, the known-background-only hypothesis has a mere likelihood of 0.19\% as compared to the hypothesis of a WIMP plus background signal \cite{cdms3events}. If indeed the events are due to WIMP-induced nuclear recoil, the resulting best-fit WIMP mass is 8.6 GeV, with a WIMP-nucleon scattering cross section of $\sigma_{\rm SI}=1.9\times10^{-41}\ {\rm cm}^2$ \cite{cdms3events}. 

The picture is complicated by the fact that the resulting best-fit region for the CDMS II results, as well as those for the DAMA/LIBRA, CoGeNT, and CRESST-II experiments, lie almost entirely above the exclusions reported by XENON100 \cite{Aprile:2012nq}, and, very recently, LUX \cite{Lux}.  However, recent studies \cite{Hooper:2013cwa, Frandsen:2013cna} have pointed out that uncertainties in the properties of liquid Xenon, as well as in the local distribution of dark matter, may be able to relieve some of this tension and bring light dark matter in the range suggested by CDMS II into better agreement with the Xenon-based experiments.  While this is a crucial point for establishing the validity of the 10 GeV WIMP hypothesis, in this study, we will largely ignore this apparent discrepancy (as well as those potentially between the CDMS II and DAMA/LIBRA, CoGeNT, and CRESST-II ``signals"), instead focusing only on achieving light WIMPs with large enough cross-sections to explain the CDMS II events, since this requirement in and of itself is often very difficult to achieve in general supersymmetric models, as we discuss below.  


If the 10 GeV WIMP scenario is taken at face value, and if one insists on requiring a WIMP with a thermal relic density in accord with the universal dark matter density, generic supersymmetric neutralinos are not natural dark matter candidates, at least in the minimal supersymmetric extension of the Standard Model (MSSM). In particular, (i) light neutralinos in the 10 GeV mass range tend to feature an excessively large thermal relic density, unless very special circumstances arise, and (ii) the typical neutralino-nucleon scattering cross section tends to be much lower than the range preferred by the direct detection experiments (see e.g. \cite{Arbey:2012na,Kuflik:2010ah, Feldman:2010ke, Boehm:2013qva} for further discussion).  For example, in the MSSM, sizable spin-independent neutralino interactions with SM fermions requires a large Higgsino fraction in the lightest supersymmetric particle (LSP), which is in tension with the lower bound from LEP on light charginos.  Also, reducing the MSSM neutralino relic density to an acceptable level in this case typically requires light sfermions to mediate the annihilation, which are generally excluded by SUSY searches at LEP and the LHC \cite{Calibbi:2013poa}, although there may yet be room for light selectrons to do the job in certain cases \cite{Boehm:2013qva}.  One exception that has been recently considered is the scenario put forth in Ref.~\cite{Arbey:2013aba}, which exemplifies the degree of difficulty involved in finding viable MSSM models with a neutralino matching the CDMS II preferred mass and cross section. This model includes a finely-tuned ultra-light right-handed sbottom, which barely evades LEP searches and constraints from flavor and precision electroweak physics, and where sbottom coannihilation suppresses the otherwise excessive relic density. This region of parameter space presents difficulties for the calculation of the neutralino-nucleon cross section \cite{Gondolo:2013wwa}, and the precise mechanism that drives the cross section to the large values needed to fit the experimental findings is somewhat unclear.



Going beyond the minimal supersymmetric scenario, the difficulties mentioned above can be alleviated by considering light Higgs states, including light CP-even Higgs bosons to mediate the spin-independent scattering of neutralinos and nucleons \cite{Bae:2010hr}.  Light scalars can arise by, e.g., adding a singlet to the superpotential of the MSSM, as in the so-called next-to-MSSM (NMSSM).  This possibility was first considered in connection with the direct detection of light neutralinos, to our knowledge, in Ref.~\cite{Gunion:2010dy}, which concluded that a $5-10$ GeV neutralino with a large spin-independent elastic scattering cross-section can be accommodated in the NMSSM provided a light enough trio of Higgs bosons appear in the theory.  Subsequently, Ref.~\cite{DLH} showed that large cross-sections $\gtrsim 10^{-41}$ cm$^2$ could be obtained with a very light ($\lesssim 5$ GeV), nearly pure singlet-like Higgs which mediates the spin-independent neutralino-nucleon scattering.  Going beyond the $\mathbb{Z}_3$-symmetric NMSSM, Ref.~\cite{Belikov:2010yi} showed that slightly heavier Higgs bosons could also accomplish this feat.  All of these scenarios additionally required a light pseudoscalar to efficiently annihilate the neutralinos in the early universe, if one insisted (as we shall do here) on a thermal relic density matching the observed universal dark matter density (for an example in which the light CP-even Higgs mediates the annihilation instead, see Ref.~\cite{Cao:2011re}).  Several other studies also focused on neutralinos with large spin-independent scattering cross-sections using full numerical scans of the NMSSM (e.g. Refs.~\cite{Cao:2010fi, Vasquez:2010ru, AlbornozVasquez:2011js}).  The aforementioned NMSSM analyses were all performed prior to the 126 GeV Higgs discovery. 

Requiring a 126 GeV SM-like Higgs alters the picture significantly \cite{Vasquez:2012hn}  (also see e.g. Ref.~\cite{godbolepaper} for a new recent analysis of light MSSM neutralinos especially in connection with effects on the Higgs sector).  In particular, light degrees of freedom can cause substantial deviations from the couplings and partial widths predicted for the Standard Model Higgs which have not been observed thus far by the ATLAS and CMS experiments at the LHC \cite{Atlas_couplings, CMS_couplings}.  Also, one of the virtues of the NMSSM is that it can provide a significant tree-level contribution to the SM-like Higgs mass, which begs the question of whether the light NMSSM neutralino scenario can be successfully realized in these regions. Additionally, the CDMS results extend the best fit region for spin-independent neutralino-nucleon scattering cross-sections down to $\sigma_{\rm SI}\gtrsim 10^{-42}$ cm$^2$, significantly lower than before, and extend the $1\sigma$ contours for the neutralino mass out to about 15 GeV, potentially re-opening portions of parameter space that were previously disfavored.  Consequently, it seems both timely and important to investigate the regions of the NMSSM producing light neutralinos compatible with the recent results from CDMS and the LHC Higgs discovery. 

In this study, we consider regions of the NMSSM in which light neutralinos \footnote{Note that it may also be possible to explain the CDMS signal via non-neutralino SUSY dark matter \cite{Choi:2013fva}, however we do not consider this possibility here.} can have a large spin-independent elastic scattering cross-section off of nucleons due to the exchange of a very light singlet-like CP-even Higgs boson.  Since the couplings of the 126 GeV Higgs are so far very SM-like \cite{Atlas_couplings, CMS_couplings}, scenarios with more than one light Higgs (as in some of the cases in Ref.~\cite{Gunion:2010dy}) are difficult to reconcile with the 126 GeV Higgs \cite{Vasquez:2012hn}, and so we do not consider them here.   We focus on two different possibilities for DM annihilation in the early universe:
\begin{enumerate}
\item \ul{\textbf{The Light CP-even Scenario:} Both the spin-independent neutralino-nucleon elastic scattering and neutralino pair annihilation rate in the early universe are mediated by the exchange of a very light ($\lesssim m_{\chi}$) singlet-like Higgs boson}.  The latter can be efficient enough if the annihilation is primarily into light CP-even Higgses, since the triple-Higgs self-coupling can be large. The most stringent constraints on this scenario arise from rare $B$ and $\Upsilon$ meson decays, as well as from decays of the Standard Model-like Higgs into light CP-even Higgses and neutralinos.  However, the corresponding branching fractions for the latter can be reduced by taking advantages of cancellations in the relevant triple-Higgs coupling and Higgs-neutralino couplings.  In this way, the SM-like Higgs can be brought into agreement with the Higgs-like particle observed at the LHC. Meson decay constraints can be substantially alleviated provided the light CP-even Higgs mass is larger than the $B$ mass, so that on-shell decays through a light scalar are prohibited; the $\Upsilon$ decay constraints are also substantially less stringent in this regime.  Note that the neutralino pair-annihilation cross-section in this scenario is $p$-wave suppressed, and hence too small to be probed by current indirect detection efforts.

\item  \ul{\textbf{The Light CP-even/CP-odd Scenario:} The elastic scattering cross-section is mediated by the exchange of a light scalar, while the neutralino pair annihilation in the early universe is mediated by a light ($\lesssim 30$ GeV) singlet-like pseudoscalar}.  Light pseudoscalars can arise in the Peccei-Quinn (PQ) or $R$-symmetric limit of the NMSSM, as well as through accidental cancellations between parameters, however we find that this setup is most readily realized near the small-$\lambda$ PQ-symmetric regime.  The couplings of the SM-like Higgs to the light CP-even/odd Higgs states and neutralinos are generically small in this case, hence softening the constraints from invisible decays and partial widths of the 126 GeV Higgs.  However, since one parameter, $\kappa$, governs both the lightest neutralino mass its spin-independent scattering cross-section, it is typically more difficult, although still possible, to obtain a LSP compatible with both CDMS II and Higgs constraints.  Rare $B$ and $\Upsilon$ decays again typically require the light scalar to have a mass larger than about 5 GeV.  Although the primary DM annihilation channel is an $s$-wave process, prospects for indirect detection are still limited at zero temperature, since one must generally sit off the resonance peak to obtain the correct relic density.
\end{enumerate}

We focus in the present study on analytical arguments to provide insight into the viable regions of parameter space.  LEP, LHC, and flavor physics constraints generally dictate that there cannot be many other degrees of freedom with significant couplings to the Standard Model (if any) with masses below around 100 GeV.  This suggests that most of the NMSSM spectrum can be decoupled from the problem, allowing us to investigate the scenario by varying relatively few quantities and without many assumptions about the rest of the spectrum.  In fact, our results here can be extended and applied to more general, non-supersymmetric models with a light Majorana fermion dark matter candidate, a light scalar and/or light pseudoscalar along the lines of e.g. Ref.~\cite{Cotta:2013jna}.  We encourage the Reader to bear this mind in the interpretation of our work.

The remainder of this study is organized as follows.  In Sec.~\ref{sec:model}, we briefly outline the Higgs and neutralino sectors of the NMSSM and the requirement of a 126 GeV Higgs on the model.  In Sec.~\ref{sec:light}, we discuss the requirement of light Higgs scalars and pseudoscalars to obtain large spin-independent neutralino-nucleon elastic scattering cross-sections and the correct relic abundance of dark matter, and describe the various other constraints on light Higgs states from colliders and flavor physics.  Sections.~\ref{sec:scalar} and \ref{sec:pseudo} comprise our analysis of the Light CP-even and Light CP-even/CP-odd cases, respectively.  We elucidate the parameter space compatible with both scenarios, providing benchmarks and commenting on the prospects for future discovery in both cases.  We discuss and conclude in Sec.~\ref{sec:conc}. 




\section{The NMSSM Higgs and Neutralino Sectors}\label{sec:model}

We begin by briefly outlining our conventions for the Higgs and neutralino sectors of the NMSSM.  We follow here the discussion of Ref.~\cite{Ellwanger:2009dp}, to which we refer the reader for a more detailed account of the model under consideration.

\subsection{The Model}
We consider the scale-invariant NMSSM, endowed with a $\mathbb{Z}_3$ symmetry prohibiting dimensionful terms in the superpotential. The latter is given by
\begin{equation} W=W_{\rm MSSM}|_{\mu=0}+\lambda \widehat{S}\widehat{H}_u\widehat{H}+\frac{\kappa}{3}\widehat{S}^3,
\end{equation}
where hatted quantities represent the chiral superfields $\widehat{H}_u=\left(\widehat{H}_u^+,\widehat{H}_u^0\right)$,  $\widehat{H}_d=\left(\widehat{H}_d^0,\widehat{H}_d^-\right)$ and where $\widehat{S}$ is a gauge singlet.  The soft supersymmetry breaking part of the Lagrangian is given by
\begin{equation}
\Delta V_{soft}=m^2_{H_u} \left|H_u\right|^2+m_{H_d}^2\left|H_d\right|^2+m_{S}^2\left|S\right|^2+\lambda A_{\lambda}H_u H_d S + \frac{1}{3}\kappa A_{\kappa}S^3.
\end{equation}
The tree-level potential relevant for the Higgs sector is given by
\begin{equation}
\begin{aligned}
V=&\frac{g^2}{4}\left(\left|H_u^0\right|^2+\left|H_u^+\right|^2-\left|H_d^0\right|^2-\left| H_d^-\right|^2\right)^2+\frac{g_2^2}{2}\left|H_u^+H_d^{0*}+H_u^0H_d^{-*}\right|^2 +\Delta V_{\rm soft} +\sum_i \left|F_i\right|^2,\\
\end{aligned}
\end{equation}
where $g^2\equiv (g_1^2+g_2^2)/2$, $g_1$ and $g_2$ denote the $U(1)$ and $SU(2)$ gauge couplings, respectively, and the sum over the $F$--terms is over $H_{u,d}^0$, $S$, with $F\equiv\partial W/\partial \phi_i$.  The neutral scalar fields can be expanded around their vacuum expectation values (vevs) as follows \cite{Agashe:2012zq}:
\begin{align}
H^0_u&=v_u+\frac{1}{\sqrt{2}}\left[\left(h_v^0+iG^0\right)\sin\beta+\left(H_v^0+i A_v^0\right)\cos\beta\right]\\
H^0_d&=v_d+\frac{1}{\sqrt{2}}\left[\left(h_v^0-iG^0\right)\cos\beta-\left(H_v^0-i A_v^0\right)\sin\beta\right]\\
S&=v_s+\frac{1}{\sqrt{2}}\left(h_s^0+iA_s^0\right).
\end{align}
In this basis it is easy to see which states couple linearly to the $W$ and $Z$.  In fact, the $h_v^0$ weak eigenstate couples at tree level to SM gauge bosons with couplings identical to that of the Standard Model Higgs.  Meanwhile, $H_v^0$ carries no tree-level couplings to $W$, $Z$.  The above expansion yields tree-level mass matrices for the CP-even and CP-odd states, after absorbing the Goldstone mode $G_0$ into the longitudinal polarization of the $Z$ boson and minimizing $V$ to eliminate the soft masses\footnote{Note that when we perform our numerical analysis we minimize the full 1-loop effective potential.} $m_{H_u}^2$, $m_{H_d}^2$, and $m_S^2$.  In the $\left(h_v^0, H_v^0, h_s^0\right)$ and $\left(A_v^0,A_s^0\right)$ bases, the mass matrices read: 
\begin{equation}
\label{eq:MS}
\mathcal{M}_S^2=\hspace{-.1cm}\left(\hspace{-.1cm}\begin{array}{ccc}\hspace{-.05cm}m_Z^2\cos^2 2\beta + \lambda^2 v^2 \sin^2 2\beta&\left(\lambda^2 v^2-m_Z^2\right)\sin 2\beta \cot2\beta& \hspace{-.1cm}2\lambda v \mu-v\sin 2\beta\left(2\kappa \mu+\lambda A_{\lambda}\right) \\  .&\hspace{-.3cm} \left(m_Z^2-\lambda^2 v^2\right) \sin^2 2\beta+\frac{2\kappa \mu^2+\lambda \mu A_{\lambda}}{\lambda \sin2\beta}&-\left(\kappa \mu+\lambda A_{\lambda}\right)\sin2\beta \cos 2\beta\\.&.&\hspace{-.1cm} \frac{4 \kappa^2 \mu^2+\kappa \mu A_{\kappa}}{\lambda^2}+\frac{\lambda v^2}{2\mu}A_{\lambda}\lambda \sin2\beta \end{array}\hspace{-.1cm} \right) 
\end{equation}
for the CP-even sector and 
\begin{equation}
\label{eq:MA}
\mathcal{M}_A^2=\left(\begin{array}{cc} \frac{2\mu}{\lambda \sin 2\beta} \left(\lambda A_{\lambda}+\kappa \mu\right)& \hspace{0.3cm} \lambda v \left(A_{\lambda}-\frac{2\kappa \mu}{\lambda}\right) \\ .&\hspace{0.3cm} \frac{\lambda v^2 \left(\lambda A_{\lambda}+4\kappa \mu\right)\sin2\beta}{2 \mu}-\frac{3\kappa A_{\kappa}\mu}{\lambda}  \end{array}\right)
\end{equation}
for the pseudoscalar sector.  It will also be useful to define the diagonalizing matrix $S$ given by 
\begin{equation}
\label{eq:S}
S^T R^T \mathcal{M}_{S}^2 R S =\operatorname{diag}\left(m_{h_1}^2,m_{h_2}^2,m_{h_3}^2\right)
\end{equation}
where 
\begin{equation}
R\equiv \left(
\begin{array}{c c c}
\cos\beta&-\sin \beta&0\\
\sin\beta&\cos \beta&0\\
0&0&1
\end{array}
\right)
\end{equation}
rotates the upper left portion of $\mathcal{M}_S^2$ by the angle $\beta$ (this rotation brings the mass matrix into the more conventional basis that enters into the Feynman rules in Ref.~\cite{Ellwanger:2009dp}).  The eigenstates of the CP-even mass matrix are denoted as $h_i$, $i=1,2,3$ (ordered in mass from lightest to heaviest), and likewise with the CP-odd Higgs mass eigenstates, $a_i$, $i=1,2$, which result from diagonalizing $\mathcal{M}_A^2$ by the matrix $P'$.  The matrix $P$ that enters the Feynman rules for the pseudoscalar couplings is actually that which diagonalizes the $3\times3$ mass matrix in the weak eigenstate basis, related to $P'$ via
\begin{equation}
P_{i1}=\sin\beta P_{i1}' ,\qquad P_{i2}=\cos\beta P_{i1}',\qquad P_{i3}=P_{i2}'.
\end{equation} 
Throughout our analysis, $h_2$ will correspond to the 126 GeV SM-like Higgs, while $h_1$ ($a_1$) will correspond to a light, singlet-like scalar (pseudoscalar).  Note that there are also charged Higgs bosons with masses set by the mass scale of $h_3$, however these states will typically be heavy in our scenario and thus will not be relevant for the phenomenology discussed here. 

Since we are especially concerned with the dark matter phenomenology of the model, it is important to review the neutralino mass matrix as well.  Let us denote the $U(1)_Y$ and neutral $SU(2)$ gauginos as $\widetilde{B}$, $\widetilde{W}$, respectively, the $H_{u,d}^0$ Higgsinos as $\widetilde{H}_{u,d}$, and the singlino as $\widetilde{S}$.  Then in the basis $\left(-i \widetilde{B},-i\widetilde{W}, \widetilde{H}_d^0,\widetilde{H}_u^0,\widetilde{S}\right)$, the neutralino mass matrix is given by 
\begin{equation} 
\label{eq:neutralino} 
\mathcal{M}_{\chi^0}=\left(\begin{array}{ccccc}M_1 \hspace{.3cm}& 0\hspace{.3cm}&-\frac{g_1v\cos \beta}{\sqrt{2}}&\frac{g_1v\sin\beta}{\sqrt{2}}\hspace{.3cm}&0\\ .&M_2&\frac{g_2v\cos\beta}{\sqrt{2}}&-\frac{g_2v\sin\beta}{\sqrt{2}}&0\\.&.&0&-\mu&-\lambda v\sin\beta\\.&.&.&0&-\lambda v\cos\beta \\.&.&.&.&2\kappa \mu/\lambda \end{array} \right). 
\end{equation} 

The above expression is diagonalized by the matrix $N_{ij}$ and the resulting lightest neutralino composition will be given in terms of the components of $N_{ij}$ as
\begin{equation}
\chi_1^0=N_{11} \widetilde{B}+N_{12} \widetilde{W} +N_{13} \widetilde{H}_d^0 + N_{14}\widetilde{H}_u^0+N_{15} \widetilde{S}.
\end{equation}
As we shall see in the following sections, we will typically be concerned here with bino- and singlino-like lightest neutralinos.

\subsection{A 126 GeV Higgs}

The tree-level SM-like Higgs mass is determined by diagonalizing Eq.~(\ref{eq:MS}).  From the upper left diagonal entry in $\mathcal{M}_S^2$, we see that the Higgs with SM-like couplings to gauge bosons has a tree-level upper bound, in the NMSSM, of 
\begin{equation} \label{eq:bound}
m_{h}^2\leq m_Z^2\cos^2 2\beta + \lambda^2 v^2 \sin^2 2\beta.
\end{equation}
For large $\lambda$ and $\tan\beta\sim1$, the expression above indicates that the tree-level mass can be large enough to accommodate $m_h\simeq 126$ GeV without requiring sizable quantum corrections, in contrast to e.g. the MSSM \cite{Feng:2013tvd}.

It has been long appreciated that quantum corrections to the Higgs masses are quite significant \cite{Gunion:1984yn} and must be taken into account in any reliable calculation of the spectrum, decay rates, cross-sections, etc.  These corrections can be addressed compactly by considering the effective action for the various Higgs fields, given through a loop expansion by
\begin{equation}
S_{\rm eff}=\int d^4x \left[\sum_{n=0}^{\infty} \left(Z_i^{n} D_{\mu} \phi_i^{\dagger} D^{\mu}\phi_i -V_n (\phi_i) \right)\right],
\end{equation}
where $\phi_i$ denote the various Higgs bosons and $Z_i^{n}$ are wave function renormalization factors. In the above expression, $V_n$ is the effective potential which, at one loop in the $\overline{DR}$ scheme is given by \cite{ColemanWeinberg}
\begin{equation}
V_1=\frac{1}{64 \pi^2}\operatorname{STr} M(\phi)^4 \left[\log\left(\frac{M(\phi)^2}{Q^2}\right)-\frac{3}{2}\right],
\end{equation}
with $M(\phi)$ denoting the field-dependent (tree-level) masses of all particles in the effective theory below the renormalization scale $Q$.  We detail which quantum corrections we take into account (and to what order) when calculating the various quantities in the following sections.

The largest 1-loop corrections to $m_h$ typically arise from the (s)top sector.  The (s)top 1-loop contribution to the Higgs mass is maximized for large $M_{\tilde{t}}\equiv \sqrt{m_{\tilde{t}_1}m_{\tilde{t}_2}}$ and trilinear coupling $A_t$ such that
\begin{equation} \label{eq:mix}
A_t=\sqrt{6}M_{\tilde{t}}+\mu \cot\beta.
\end{equation}
This setup is known as the ``maximal mixing scenario".  On the other hand, the most significant contribution to the singlet-like CP-even mass eigenvalue is from neutralinos and charginos running in the loop.  Full expressions for these various contributions to the mixing matrices can be found in Ref.~\cite{Ellwanger:2009dp}, whose conventions we follow throughout this work.
 
\section{Light Neutralinos and Large Cross-Sections via Light Higgses} \label{sec:light}

We are interested in neutralinos with a large enough $\sigma_{\rm SI}$ to explain the events observed by the CDMS experiment \cite{cdms3events}.  To see what this requirement implies for the spectrum, we can consider the following low-energy effective four-fermi interaction Lagrangian, which governs the spin-independent scattering in the NMSSM:
\begin{equation}
\mathcal{L}_{\rm eff}\supset \sum_{i} a_{q_i} \chi_1^0 \chi_1^0 q_i \bar{q}_i.
\end{equation}
where the sum runs over all quark flavors.  Terms with pseudoscalar couplings contribute only to the spin-dependent cross-section and are omitted, as are terms that are velocity- or momentum transfer - suppressed (such as those arising from vector exchange).  The cross-section for the spin-independent interaction of a neutralino with a proton ($p$) or neutron ($n$) is then given by \cite{Ellis:2000ds}
\begin{equation}
\label{eq:SI}
\sigma_{\rm SI}^{p,n} =\frac{4 m_{\chi}^2 m_{p,n}^4}{\pi \left(m_{\chi}+m_{p,n}\right)^2}\left[\sum_{i=u,d,s} \frac{a_{q_i} f_{q_i}^{p,n}}{m_{q_i}}\right]^2,
\end{equation}
where the sum is over the appreciable quark constituents of the proton and neutron and the hadronic matrix elements $f_{q_i}^{p,n}$, given by
\begin{equation}
m_{p,n} f_{q_i}^{p,n}\equiv \langle p,n| m_{q_i}q_i \bar{q}_i| p,n \rangle,
\end{equation}
specify the quark content\footnote{There is a sizable uncertainty in the strange quark content of the proton, and hence in $f_s^p$, which affects the computation of $\sigma_{\rm SI}^p$.  We assume $\sigma_0=35$ MeV and $\sigma_{\pi N}=45$ MeV throughout our analysis.} of the nucleons. The couplings $a_{q_i}$ for neutralinos to up- and down- type quarks through Higgs mediators are given by
\begin{equation}
\label{eq:a_q}
\begin{aligned} a_{q_i}&=\frac{m_{q_i}}{\sqrt{2} v \sin \beta} \sum_{j=1}^3 \frac{g_{h_j \chi \chi} S_{j2}}{m_{h_j}^2}, \qquad q_{i}=u,s,t \\
a_{q_i}&=\frac{m_{q_i}}{\sqrt{2} v \cos \beta}  \sum_{j=1}^3 \frac{g_{h_j \chi \chi} S_{j1}}{m_{h_j}^2}, \qquad q_{i}=d,c,b
\end{aligned}
\end{equation}
where $m_{q_i}$, $m_{h_i}$ are the various quark and Higgs (pole) masses, $v\simeq 174$ GeV, and $g_{h_j\chi\chi}$ is the relevant coupling between $h_j$ and the lightest neutralino $\chi_1^0$. $S_{ij}$ is defined in Eq.~(\ref{eq:S}).  Note that when we write $\sigma_{\rm SI}$ without the $p$ or $n$ superscript, we mean the average of the two quantities.

From Eq.~(\ref{eq:SI}) it is straightforward to see why it is generally difficult to achieve a large $\sigma_{\rm SI}$ for light neutralinos in the MSSM.  First of all, the Higgsino and wino components of $\chi_1^0$, which govern $g_{h_j \chi \chi}$ as well as the neutralino couplings to gauge bosons, cannot be very large for a sub-10 GeV neutralino, since $\mu$ and $M_2$ are both constrained to be $\gtrsim 100$ GeV by LEP and because a large Higgsino component leads to too large an invisible branching fraction for the $Z$ boson. Secondly, for the case in which the scattering is mediated by the exchange of a CP-even Higgs boson $h_j$, the cross-section scales as $\sim 1/m_{h_j}^4$ and is therefore suppressed in the MSSM.  An estimate for the cross-section in the most optimistic MSSM case (with large $\tan\beta$, significant $N_{13}$, and minimal Higgs mixing) yields \cite{Belikov:2010yi}
\begin{equation}
\sigma_{\rm SI}^{\rm MSSM} \approx 1.8\times 10^{-41} {\rm cm^2} \left(\frac{N_{13}^2}{0.103}\right) \left(\frac{\tan\beta}{50}\right)^2\left(\frac{90 \hspace{.1cm} {\rm GeV}}{m_h}\right)^4 \left(\frac{S_{11}}{1}\right)^4\lesssim 5 \times 10^{-42}\hspace{.1cm} {\rm cm}^2,
\end{equation} where the inequality follows for $m_{h}=126$ GeV and all other ratios set to unity.  This estimate is over-optimistic, not taking into account constraints on e.g. the Higgs properties and LHC neutralino/chargino searches. Consequently, obtaining light neutralinos with $\sigma_{\rm SI}\gtrsim 10^{-42}$ cm$^2$ has been shown to be difficult or impossible in SUSY models with minimal field content \cite{  Kuflik:2010ah, Feldman:2010ke}. 

 
Even without requiring a large $\sigma_{\rm SI}$, there is another important reason that sub-15 GeV MSSM neutralino DM is difficult to come by.  In order for $\chi_1^0$ to be a viable dark matter candidate, the neutralinos must be able to annihilate efficiently in the early universe.  The WMAP and recent PLANCK results bound the thermal relic density to fall within the range \cite{PLANCK, WMAP9}
\begin{equation}
\label{eq:relden}
0.091\leq \Omega h^2\leq 0.138,
\end{equation}
where $h$ is the local Hubble expansion parameter in units of 100 km/s/Mpc.  These bounds correspond to the $2\sigma$ limits from the WMAP 9-year data with $10\%$ theoretical uncertainty, which also encompasses the range suggested by PLANCK.  For moderately heavy neutralinos, one can have a ``well-tempered" neutralino \cite{Masiero:2004ft, ArkaniHamed:2006mb} where the Higgsino, wino, and bino components of the neutralino precisely balance to give the correct thermal relic abundance.  However, for light neutralinos this is difficult since LEP limits on light charginos dictate that $\operatorname{Min}\left\{M_2,\mu\right\}\gtrsim 100$ GeV \cite{LEP_chargino}.  Also, for such light WIMPs, LEP limits on light superpartners significantly constrain sfermion mediation or co-annihilation as possible mechanisms to dilute $\Omega h^2$ \cite{Calibbi:2013poa}.  For this reason, light neutralinos in the MSSM are difficult to come by, even without requiring a large $\sigma_{\rm SI}$ \cite{Arbey:2012na, Calibbi:2013poa, godbolepaper}.

In the NMSSM, the presence of a new scalar and/or pseudoscalar, as well as the singlino contribution to the neutralino sector, can rescue light dark matter.  In particular, if the CP-even or CP-odd Higgs bosons are light enough, they can mediate neutralino pair-annihilation in the early universe.  Meanwhile, the exchange of a light scalar can contribute to the elastic scattering cross-section and, for a light enough singlet-like $h_1$, the $1/m_{h_1}^4$ suppression can be overcome to provide $\sigma_{\rm SI}\sim10^{-42}$--$10^{-41}$ cm$^2$, as pointed out e.g. in Refs.~\cite{Gunion:2010dy, DLH, Belikov:2010yi}.  As we will see in the following sections, to reach the CDMS best fit region typically requires a singlet-like Higgs with mass $m_{h_1}\lesssim 10$ GeV (this rough upper limit can be increased in going beyond the $\mathbb{Z}_3$-symmetric NMSSM as shown in Ref.~\cite{Belikov:2010yi}, however, we do not consider this case here). The lightest neutralinos can couple sizably to the light singlet-like scalar and pseudoscalar even if the latter are pure singlets, due to the singlino component in the neutralino.  Thus, light CP-even and CP-odd Higgs bosons can lend light neutralinos the necessary ingredients to be viable light dark matter candidates with large spin-independent neutralino-nucleon elastic scattering cross-sections as suggested by CDMS and with the correct thermal relic abundance.  We consider in detail both the light scalar and the light pseudoscalar annihilation scenarios, and the dependence of the relevant cross-sections on the various parameters below.



\subsection{Constraints}\label{subsec:constraints}

As expected, collider searches and flavor physics set stringent limits on the couplings of new light degrees of freedom to the Standard Model.  Here we review the key experimental constraints on the scenarios under consideration. 

\subsubsection{LEP and Tevatron Constraints}
The existence and properties of light Higgs bosons have long been constrained by searches at LEP and the Tevatron.  By considering the various decay topologies of Higgstrahlung production $e^+ e^- \rightarrow Z h_{1,2}\rightarrow(\ldots)$ as well as Higgs pair production $e^+ e^-\rightarrow h_1 h_2 \rightarrow (\ldots)$, LEP searches long ago ruled out a Standard Model-like Higgs boson below $\sim115$ GeV \cite{Barate:2003sz}.  These results, taken in conjunction with the apparent SM-like nature of the 126 GeV resonance observed at the LHC, dictate that the light CP-even Higgs in our scenario must be very singlet-like.  

Since we will be concerned with a very light $h_1$, the most constraining LEP searches are the decay channel-independent light Higgs searches from ALEPH at LEP1 \cite{ALEPH} and OPAL at LEP2 \cite{Abbiendi:2002qp}, which set limits on the coupling of $h_1$ to the SM gauge bosons.  In the basis of Eq.~(\ref{eq:MS}), this amounts to an upper bound on the $h_v^0-h_s^0$ mixing, which we consider in more detail below.  Both LEP and the Tevatron also constrain the $h_2h_1h_1$ and $h_1f\bar{f}$ couplings (where $f$ is a SM fermion) through searches for e.g. $h_2\rightarrow h_1 h_1 \rightarrow 4b, 4\tau, 2b 2\tau$ decays at LEP \cite{Schael:2006cr, Schael:2010aw} and $h_2\rightarrow h_1 h_1 \rightarrow 2\mu^{+}2\mu^-$ decays at the Tevatron \cite{Abazov:2009yi} (these searches also apply to a light pseudoscalar $a_1$).  We impose these constraints (as well as all others implemented in \texttt{HiggsBounds} \cite{HiggsBounds} and \texttt{NMSSMTools} \cite{NMSSMTools}) on our parameter space. As we will see, however, all of the aforementioned constraints are typically \textit{eo ipso} satisfied provided $h_2$ is consistent with the Higgs boson observed by CMS and ATLAS.  

\subsubsection{LHC Higgs Searches}

The discovery of a 126 GeV SM-like Higgs boson provides a whole other set of requirements on our scenario, namely that the couplings of $h_2$ (and the corresponding production cross-sections and decay rates) be in agreement with those measured by the CMS and ATLAS experiments at the LHC \cite{Atlas_couplings, CMS_couplings, Belanger:2013xza}.  By requiring $h_2$ to be SM-like and $h_1$ singlet-like, the couplings of $h_2$ to SM degrees of freedom can easily be made similar to the vanilla SM Higgs; for the scenarios we consider here, the couplings of $h_2$ to quarks and SM gauge bosons will typically be within $5\%$ of the values predicted by the Standard Model. However, the presence of a light $h_1$ and $\chi_1^0$ can cause substantial deviations in the relevant production cross-sections and result in unobserved decay properties.  This class of constraints can be compactly addressed by considering the total decay width of $h_2$, $\Gamma_{h_2}^{tot}$ \cite{Barger:2012hv} as well as the $h_2$ invisible branching fraction.   Since these quantities depend sensitively on the light CP-even/odd Higgs bosons and lightest neutralino(s), they serve as powerful discriminators in the NMSSM regions of interest and will function as our primary check on $h_2$ against the observed 126 GeV Higgs boson (although we investigate the other Higgs reduced couplings as well).

Explicitly, the partial width of $h_2$ decaying to generic SM final states is given by 
\begin{equation}
\Gamma_{h_2}^{vis.}=\sum_{Y\bar{Y}} \Gamma\left(h_2\rightarrow Y \bar{Y}\right)=\sum_{Y\bar{Y}}\kappa_Y^2 \Gamma\left(h_{SM}\rightarrow Y \bar{Y}\right),
\end{equation}
where $Y$ denotes the various SM final states $Y \bar{Y}=b\bar{b}, WW^*,\ldots$ and $\kappa_Y$ is the $h_2$ reduced coupling to $Y\bar{Y}$, given by $\kappa_Y^2\equiv \left|c_Y\right|^2/\left|c_Y^{\rm SM}\right|^2=\Gamma(h_2\to Y\bar{Y})/\Gamma(h_{\rm SM}\to Y\bar{Y})$.  Here $c_Y$, $c_Y^{\rm SM}$ are the effective $h_2$, $h_{\rm SM}$ couplings to $Y\bar{Y}$, entering the 1PI effective Higgs interaction Lagrangian 
\begin{equation} \label{eq:Leff}
\begin{aligned}
\mathcal{L}_{eff} =& c_V \frac{2 m_W^2}{v}h W_{\mu}^+W_{\mu}^- + c_V \frac{m_Z^2}{v}h Z_{\mu}Z_{\mu}-\sum_f c_f \frac{m_f}{v}h f \bar{f}\\
&+c_g \frac{\alpha_s}{12 \pi v}h G_{\mu \nu}^a G_{\mu \nu}^a + c_{\gamma} \frac{\alpha}{\pi v} h A_{\mu \nu}A_{\mu \nu}
\end{aligned}
\end{equation}
at the scale $m_{h}\simeq 125$ GeV \cite{Carmi:2012in} where $h=h_2,h_{\rm SM}$ and $f$ denotes the relevant SM fermions, not including the top quark; the top is integrated out in Eq.~(\ref{eq:Leff}) which gives rise to the dimension-5 couplings of $h$ to gluons and photons in the second line (see Ref.~\cite{Carmi:2012in} for detailed expressions for these couplings, including contributions from new physics, which, in our case, primarily comprises diagrams with a chargino running in the loop).  In the SM, at tree-level, $c^{\rm SM}_V=c^{\rm SM}_f=1$ and so $\kappa_Y\simeq c_Y$ for these states.
The total width of $h_2$ additionally includes contributions from invisible decay processes,
\begin{equation}
\Gamma_{h_2}^{tot}=\Gamma_{h_2}^{vis.}+\Gamma_{h_2}^{{\rm invis}.}.
\end{equation}
Using the narrow width approximation, the agreement with the Standard Model prediction for the various production and decay channels can be quantified by considering the ratio
\begin{equation} \label{eq:width}
\mu_{XY}\equiv \frac{\left(\sigma\cdot BR\right)\left(X \bar{X}\rightarrow h_2\rightarrow Y\bar{Y}\right)}{\left(\sigma\cdot BR\right)\left(X \bar{X}\rightarrow h_{SM}\rightarrow Y\bar{Y}\right)}=\frac{\kappa^2_{X} \kappa^2_{Y}\Gamma_{h_{SM}}^{tot}}{\Gamma_{h_2}^{tot}},
\end{equation}
where $X$ is the initial state relevant for the Higgs production process and where $\kappa_{X}$ is defined analogously to $\kappa_{Y}$. 

From Eq.~(\ref{eq:width}) above, it is clear that the observable $\mu_{XY}$ constrains both the reduced couplings $\kappa_{X,Y}$ and the total width $\Gamma_{h_2}^{tot}$.  We check the reduced couplings in our scenario against the 95\% C.L. ellipses for these quantities obtained from the global fit performed in Ref.~\cite{Belanger:2013xza}.  These constraints should be taken with a grain of salt, however.  Since $\kappa_{g,\gamma}$ are induced at one-loop level and beyond, they generally depend quite sensitively on the SUSY spectrum beyond the requirements for light DM.  For example, the lightest chargino contribution to $c_{\gamma}$ depends sensitively on $\mu$ and $M_2$, while the value and sign of $M_2$ does not otherwise significantly affect the DM or Higgs phenomenology considered here. Thus, $\kappa_{g,\gamma}$ can be varied quite substantially while leaving the light DM scenario in tact.  

A more robust constraint will be provided by the $h_2$ total width and invisible branching fraction, as these quantities depend directly on the light neutralino and Higgs spectrum.  Even if all of the couplings of $h_2$ to SM degrees of freedom are close to those of the Standard Model Higgs (as they will be in the scenarios we consider), sizable deviations from the expected SM production cross-sections can arise if the total width of $h_2$ differs significantly from the predicted SM value $\Gamma_{h_{SM}}^{tot}\approx 4.1$ MeV \cite{Dobrescu:2012td}.  Of course, if the reduced couplings $\kappa_X$, $\kappa_Y$ were significantly larger than unity,  this could in principle balance out an enhanced total width, however we do not find this to be the case here.  Consequently, both exotic visible and invisible decays of the SM-like Higgs are tightly constrained by the observed signal strengths for the various SM channels \footnote{For the cases considered here, since the couplings of $h_2$ to the $SU(2)$ gauge bosons are typically very close to or slightly below unity, the invisible branching fraction is more constrained by fits to the Higgs couplings than by direct searches for e.g. $ZH\rightarrow l^+l^-+\slashed{E}_T$.  See e.g. Ref.~\cite{Belanger:2013xza}.}, since they contribute to the Higgs total width \cite{Bai:2011wz, Belanger:2013kya}.  Inferred bounds on the total and invisible widths depend on the various $\kappa_{X,Y}$.   The global fit analysis performed in Ref.~\cite{Belanger:2013xza} treating the SM-like Higgs couplings to up- and down-type fermions, $SU(2)$ gauge bosons, photons, and gluons, as free parameters suggests $\Gamma_{h_2}^{tot}/\Gamma_{h_{SM}}^{tot} \lesssim 2$ and $BR(h_2\rightarrow {\rm invis}.)\lesssim 36\%$, both at 95$\%$ C.L..  In the case where all couplings are as in the SM, the bound is stronger still: $\Gamma_{h_2}^{tot}/\Gamma_{h_{SM}}^{tot} \lesssim 1.3$, $BR(h_2\rightarrow {\rm invis}.)\lesssim 20\%$ \cite{Belanger:2013xza, Falkowski:2013dza}.  

In the two cases we investigate here, the largest new contributions to $\Gamma_{h_2}^{tot}$ are from $h_2\rightarrow h_1 h_1$ and $h_2\rightarrow \chi_1^0 \chi_1^0$ decays, with the latter comprising the main contribution to the $h_2$ invisible branching fraction  (for our purposes, the decay into $a_1$ pairs is relevant only near the small-$\lambda$ PQ-symmetry limit of the NMSSM, where the coupling of $h_2$ to $a_1$ is suppressed).  In order for $h_2$ to be in agreement with the limits outlined above, both the $h_2h_1h_1$ and $h_2 \chi_1^0 \chi_1^0$ couplings must be small.  We discuss how one might achieve this in the following Sections.   

\subsubsection{$B$-physics}

Rare $B$ decays add highly non-trivial constraints on light Higgs scalars and pseudoscalars.  When a scalar or pseudoscalar has a mass below the $B$ meson mass, on shell decays of the $b$ quark to $h$, $a \rightarrow \mu^+\mu^-$ can give rise to a signal in both inclusive ($B\rightarrow X_s \mu^+ \mu^-$) and exclusive (e.g. $B\rightarrow K \mu^+ \mu^-$) channels, which are highly constrained by LHCb \cite{LHCb}, Belle \cite{BELLE}, and BaBar \cite{BaBar}.  This dangerous on-shell decay depends only on the coupling of $h$ or $a$ to $b$-quarks, dictating that $S_{11}/\cos\beta \lesssim 10^{-3}$ for $m_{h_1}\lesssim 5$ GeV \cite{Batell:2009jf}.  It is very difficult to obtain such small couplings while retaining the large neutralino scattering cross-section required to explain CDMS II.  Thus, we will typically only consider $m_{h}\gtrsim 4.8$ GeV (so that $m_{h_1}\gtrsim m_B-m_K$) to avoid these constraints altogether (note that $a_1$ is always heavier than 5 GeV).  This important constraint seems to have been missed in previous work on light neutralinos in the NMSSM, as the exclusive searches are not taken into account by \texttt{NMSSMTools}.  Taking $m_{h}>4.8$ GeV significantly reduces the $1/m_h^4$ enhancement of $\sigma_{\rm SI}$, however we still find regions of parameter space that can explain the CDMS II signal.  

There are some caveats to the above statements, however.  For one, there is a large uncertainty in the branching fractions of $h_1$ when $m_{h_1}\sim 1$ GeV, due to the $f_0$ $0^+$ hadronic resonance \cite{Gunion:1989we}. As a result, $BR(h_1\rightarrow \mu^+\mu^-)$ is significantly suppressed in this region, and one might hope that a light scalar may have escaped detection by experiments probing final states with $\mu^+ \mu^-$.  However, we find that even with the most optimistic results for the reduced $BR(h_1\rightarrow \mu^+\mu^-)$ in Ref.~\cite{Gunion:1989we}, it is still quite difficult, if not impossible, to accommodate such a light scalar in a way consistent with LHCb, Belle, and BaBar results in the scenarios we consider below.

Another possible exception may arise if $h_1$ lies near the $J/\psi$ or $\psi(2S)$ resonances.  Due to the presence of these states, the LHCb, Belle, and BaBar experiments veto dimuon invariant masses in the rage $2.95\leq m_{\mu\mu} \leq 3.18$ GeV and $3.59\leq m_{\mu \mu}\leq 3.77$ GeV.  It may thus be possible in principle for $h_1$ to lie in these narrow regions and to have thereby evaded detection.  This possibility is still highly constrained by $\Upsilon$ decays and we do not dwell too much on this scenario because of the conspiracy of parameters it requires.  Still, this might still be a viable option for obtaining a large spin-independent scattering cross-section without violating current experimental constraints, and we provide a benchmark along these lines below.

In addition to prohibiting the on shell $h$-mediated $B$ decay processes, we take care to choose parameters such that the constraints from e.g. $b\rightarrow s \gamma$, $B_s\rightarrow \mu \mu$ are satisfied.  We use \texttt{NMSSMTools} to check against these constraints.\footnote{We do not require the muon $g-2$ to fall within the experimental limits.  One can bring this observable into agreement with observation by altering the details of the sfermion sector which would leave the DM and Higgs phenomenology intact, as long as $m_{h_2}=126$ GeV and the sfermions are not too light.}

\subsubsection{$\Upsilon$ Decays}

Another important set of constraints is supplied by radiative $\Upsilon(nS)$ decays, $\Upsilon\rightarrow \gamma (h_1,a_1)\rightarrow \gamma (\mu\mu,\tau\tau, gg, \hspace{.1cm} {\rm hadrons})$ \cite{ups_hadronic, ups_tau, ups_muon, ups_gluon,McKeen:2008gd, Aubert:2009cp,DLH}.  Limits on decays involving a light scalar affect the allowed coupling of $h_1$ to $b$-quarks.  This coupling must be somewhat significant in order to provide a sizable spin-independent scattering cross-section off of down-type quarks in the nucleon.  Nevertheless, the limits from $\Upsilon$ decays can be satisfied in both of the scenarios we consider.  For $m_{h_1}\sim 5$ GeV, existing experimental limits dictate that $S_{11}/\cos\beta \lesssim 0.6$ \cite{McKeen:2008gd, ups_hadronic} ($S_{11}/\cos\beta$ is the effective down-type coupling of $h_1$ to fermions), while for smaller masses the constraints are more stringent, $S_{11}/\cos\beta \lesssim 0.2$ for $m_{h_1}\lesssim 3$ GeV. Typically the most stringent constraints will arise from final states with a $\tau \tau$ pair \cite{ups_tau}.  We include these constraints on the relevant parameter space in the following Sections, computing the branching ratios using the methods outlined in Ref.~\cite{Gunion:1989we} and taking conservative choices for e.g. the QCD parameters when possible. In considering an additional light pseudoscalar, we will be primarily focused on the regime where $m_{a_1}>m_{\Upsilon}$ and so the relevant limits will typically be satisfied from the outset.

\subsubsection{Other Constraints}
Finally, there are other potential constraints on light neutralinos that are quite easily avoided or that do not significantly affect the parameter space (although we take them into account).  Collider mono-jet searches place limits on both the spin-independent and spin-dependent neutralino-nucleon elastic scattering cross-sections in an effective field theory framework \cite{ ATLAS:2012ky, Chatrchyan:2012me}.  However, for the range of parameters we consider in both the light CP-even and light CP-even/CP-odd Higgs cases, all points easily fall below the relevant bounds (see e.g. Ref.~\cite{Cotta:2013jna}).  This is because of 1) the relatively small Higgsino component in the neutralino (this is in contrast to the MSSM case \cite{Calibbi:2013poa}), 2) the fact that the light scalar mediator will always be less than twice the LSP mass, prohibiting on-shell $h_1\rightarrow \chi_1^0\chi_1^0$ decays, and 3) because the quantity $g_{a_1\chi\chi}g_{a_1 bb}$ is small in the light CP-even/CP-odd scenario.  Also, bounds on the invisible width of the $Z$-boson \cite{ALEPH:2005ab} are easily satisfied, again by virtue of the small Higgsino component in $\chi_1^0$.  

Since there is some freedom in choosing particular gaugino masses, and since $\chi_1^0$ will typically have a small Higgsino component, LEP does not place strong constraints on the other neutralino masses, but does require charginos to be heavier than $\sim 100$ GeV \cite{LEP_chargino}, which translates into a lower limit on $\mu$, $M_2$: $\operatorname{Min}\left\{\mu,M_2\right\}\gtrsim 100$ GeV.  We check against all relevant LEP constraints on associated neutralino and chargino production, as implemented in \texttt{NMSSMTools}.  Note that the LHC also sets limits on the production rates of charginos and neutralinos (the only light supersymmetric particles in our case), especially in the case of light winos.  However these constraints are alleviated by taking $M_2$ to be large, along with the sfermion masses, which are not strictly dictated in the physical setups we consider here (see e.g. Ref.~\cite{Calibbi:2013poa} for a discussion of these constraints on light MSSM neutralinos).  The Higgsinos in our scenarios will typically be light, however the corresponding LHC production cross-section for Higgsino-like charginos and neutralinos are substantially reduced relative to the corresponding rates for wino-like states. Using \texttt{MadGraph 5} \cite{Madgraph} to rescale the production cross-sections and comparing with the wino limits from ATLAS \cite{ATLAS_EW} and CMS \cite{CMS_EW}, we find roughly that taking $\mu\gtrsim 150$ GeV allows us to satisfy the relevant constraints in the cases we consider (that is, assuming a 100\% branching ratio of the Higgsino states to final states involving $\chi_1^0$, which is also conservative). The question of the neutralino and chargino production cross-sections is an interesting one and we intend to address these potential signatures in an upcoming publication.  However, since $\mu$ can be generically be larger in our scenario compared to the MSSM (given that the light Higgs scalar and/or pseudoscalar dominate the finite-temperature annihilation rate and $\sigma_{\rm SI}$), these constraints will be weaker than in the MSSM case \cite{Calibbi:2013poa} and LHC limits on the invisible branching fraction and total width of $h_2$ are expected to provide more stringent limits.

Finally, when the effective potential includes a singlet degree of freedom, the `physical' vacuum, in which electroweak symmetry is spontaneously broken, may no longer be the most energetically favorable configuration \cite{Kanehata:2010ci, Kobayashi:2012xv}.  In the scenarios we consider, this tends to happen for large values of the tri-singlet SUSY-breaking coupling $A_{\kappa}$ and/or relatively light sfermions.  Clearly such a situation is incompatible with our universe, and so we check against this constraint using \texttt{NMSSMTools}.

\section{The Light CP-even Scenario: Annihilation through a light scalar} \label{sec:scalar}

Let us first consider the case of a light neutralino accompanied by a light CP-even singlet-like Higgs boson $h_1$, the latter being responsible for the two key phenomenological aspects we are interested in: (1) mediating the direct detection neutralino-nucleon cross section as well as (2) neutralino pair-annihilation in the early universe.  We emphasize that we do not impose the requirement of a light pseudoscalar (the case with a light pseudoscalar is considered separately in Sec.~\ref{sec:pseudo} below). 

A light CP-even state contributes to the finite temperature thermally-averaged neutralino pair annihilation rate through $s$-channel $h_1$ exchange as well as  $t$- and $u$-channel neutralino exchange into $h_1h_1$ final states, provided they are kinematically allowed \footnote{If $m_{h_1}>m_{\chi}$, $h_1$-mediation can also contribute to the annihilation rate via e.g. four-body fermionic final states, however these contributions to the annihilation rate are also significantly suppressed by the single-like nature of $h_1$. We find that $h_1$ must typically be substantially lighter than $\chi_1^0$ to have efficient enough annihilation through a light scalar in the early universe.} (the cross-sections for annihilations into fermion final states are suppressed by the required singlet-like nature of $h_1$).  The $t$ and $u$ channel contributions are sub-dominant, so neutralino pair-annihilation proceeds primarily through the $s$-channel process, with the corresponding annihilation rate at $T=0$ given by
\begin{equation}
\label{eq:sigmah}
\sigma v =\frac{\left|g_{h_1 \chi \chi}\right|^2\left| g_{h_1h_1h_1}\right|^2 \left(s-4 m_{\chi}^2\right)}{64\pi s \left(s-m_{h_1}^2\right)^2}\sqrt{\frac{s-4m_h^2}{s}},
\end{equation}
where $s$ is the center-of-mass (COM) energy squared, corresponding to $4 m_{\chi}^2$ at zero temperature, and $g_{h_1 \chi \chi}$, $g_{h_1h_1h_1}$ are the couplings of $h_1$ to two $\chi^0_1$ states and the $h_1$ self-coupling, respectively (see e.g. Refs.~\cite{Drees:1992am, Ellis:2000ds} for a full expression).  As discussed in the previous section, LEP highly constrains the mixing of $h_1$ with the other CP-even states, so $h_1$ must be very singlet-like, in which case $g_{h_1 h_1 h_1}$ is given by
\begin{equation} \label{eq:ghhh}
g_{h_1 h_1 h_1} \simeq 6 \sqrt{2} \kappa^2 \mu/\lambda +2 \kappa A_{\kappa}/\sqrt{2}. 
\end{equation}
This coupling has mass dimension 1 and can be sizable provided $\kappa$ and $A_{\kappa}$ are not both too small.  Since $g_{h_1 h_1 h_1}$ must be large to allow for efficient neutralino pair-annihilation at finite temperature, moderate values of $\kappa$ will typically be required, thereby reducing the singlino component in the LSP (see Eq.~(\ref{eq:neutralino})).  In conjunction with the LEP limits on charginos, this implies that the lightest neutralino must be predominantly bino-like with a small Higgsino fraction to couple it significantly to $h_1$ (which must in turn have a small $SU(2)$ component), in which case $g_{h_1 \chi \chi}$ is given by
\begin{equation} \label{eq:ghnn}
g_{h_1 \chi \chi} \simeq \sqrt{2}\lambda N_{13}N_{14}. 
\end{equation}

The contribution of Eq.~(\ref{eq:sigmah}) to the annihilation rate is $p$-wave suppressed, and hence vanishes at $v=0$.  Consequently, indirectly detecting these neutralino annihilations through gamma-ray or charged cosmic-ray observations is not likely.  However in the early universe, the annihilation rate is given by the thermal average
 \cite{Gondolo:1990dk}:
\begin{equation} \label{eq:thermal}
\langle \sigma v\rangle_{T\neq0}=\frac{1}{8m_{\chi}^4 T K_2(m_{\chi}/T)^2}\int_{4m_{\chi}^2}^{\infty}\sigma(s)\left(s-4m_{\chi}^2\right)\sqrt{s}K_1\left(\sqrt{s}/T\right),
\end{equation}
where $K_{1,2}$ are modified Bessel functions of the first- and second-kind, respectively \cite{Gondolo:1990dk}.  At finite temperature (and velocity), the annihilation rate can thus be large enough to drive the relic density down without introducing any additional degrees of freedom, as we will see below.  

For $\chi_1^0 \chi_1^0\to h_1\to h_1 h_1$ annihilation to be kinematically allowed requires $m_{h_1}\leq m_{\chi}$.  If $m_{h_1}$ is very light, the contribution of $h_1$ exchange to $\sigma_{\rm SI}$ can be large, raising the elastic scattering cross-section to the levels suggested by CDMS II.  The relevant contribution to the cross-section is given in Eq.~(\ref{eq:SI}).

\subsection{The Parameter Space}\label{sec:choose}

To hone in on corresponding viable regions of the NMSSM parameter space, we consider the following three general requirements:
\begin{itemize}
\item A Standard Model-like second-lightest Higgs boson ($h_2$) consistent with the resonance observed at the LHC with mass $m_{h_2}\sim 126$ GeV;

\item A lightest neutralino LSP with mass $m_{\chi_1}\sim 5-15$ GeV and with a thermal relic abundance in the range dictated by WMAP and PLANCK, $0.09\lesssim \Omega h^2 \lesssim 0.14$;

\item A large spin-independent neutralino-nucleon elastic scattering cross-section $10^{-42}$ cm$^2 \lesssim \sigma_{\rm SI} \lesssim  10^{-40}$ cm$^2$ as required to explain the CDMS signal \cite{cdms3events}.  To fulfill this requirement, we impose a singlet-like lightest CP-even Higgs with mass $m_{h_1}$ in the range $m_{h_1} \lesssim m_{\chi_1}$ and consistent with constraints from the LHC, Tevatron, LEP, and flavor physics.

\end{itemize}

Let us now review how each of the three conditions affects the parameter space.  As per Eq.~(\ref{eq:bound}), the tree-level Higgs mass can be substantially higher in the NMSSM than in the MSSM (where the upper bound is $m_Z$ for large $\tan\beta$) and so a heavy Higgs can arise rather naturally in this model.  The NMSSM tree-level contribution to $m_{h_2}$ is maximized for $\tan\beta\sim 1$ and large $\lambda$.  Alternatively, large sfermion masses and mixing parameters can raise $m_{h_2}$ to the desired level as in the MSSM.  We find that the light CP-even scenario generally requires a moderate contribution from sfermion effects to raise the Higgs mass.  To see this, note that the coupling of $h_2$ to $\chi_1^0$ in our case is approximately given by
\begin{equation}
\label{eq:h2neutcoupling}
g_{h_2\chi\chi}\approx \frac{2\lambda}{\sqrt{2}}\left(S_{21} N_{14}N_{15}+S_{22}N_{13}N_{15}\right)+g_1\left(S_{21}N_{11}N_{13}-S_{22}N_{11}N_{14}\right),
\end{equation}
so $g_{h_2\chi\chi}$ is enhanced when $\lambda$ is large, since $\chi_1^0$ typically has small but non-vanishing Higgsino and singlino components.  This coupling affects the branching ratio $BR(h_2\rightarrow \chi_1^0 \chi_1^0)$ which must fall below $\sim38\%$ to be consistent with the LHC Higgs signal at $95\%$ C.L \cite{Belanger:2013xza}.  This would suggest that $\lambda$ cannot be too large.  On the other hand, the spin-independent scattering cross-section is governed by $g_{h_1\chi\chi}$, which is proportional to $\lambda$ (see Eq.~(\ref{eq:ghnn})).  Since $B$ and $\Upsilon$ physics generally dictate $m_{h_1}\gtrsim 5$ GeV as argued above, the $1/m_{h_1}^4$ enhancement of $\sigma_{\rm{SI}}$ cannot be arbitrarily large, and so significant values of $g_{h_1\chi\chi}\propto \lambda$, are required.  Thus in the light CP-even scenario there is tension between requiring a large $\sigma_{\rm SI}$ and a SM-like Higgs in agreement with observation.  Also, note that small values of $\tan\beta$ weaken the coupling of $h_2$ to down type fermions, which also tends to increase the branching ratio of $h_2$ into neutralinos and decrease $\sigma_{\rm SI}$.  On the other hand, if $\tan\beta$ is too large, constraints from radiative $\Upsilon$ decays and other flavor processes become more severe.  Thus, both $\lambda$ and $\tan \beta$ will need to fall in intermediate ranges to satisfy all constraints from experiment and simultaneously explain the CDMS II signal.  This means that moderately heavy sfermion masses will be required to raise the Higgs mass to 126 GeV, however the tuning is not egregious.  In practice we will consider $\lambda$ to be in the range $0.5\leq \lambda \leq 0.6$ and $5 \leq \tan\beta \leq 10$; these choices provide a sizable $\sigma_{\rm SI}$ and a moderate NMSSM tree-level contribution to $m_{h_2}$, while, for typical choices for the other parameters in our scan, also allow for $BR(h_2\rightarrow \chi_1^0\chi_1^0)\lesssim 35\%$.  

A similar line of reasoning informs our choice for the value of $\mu$.  In the scale-invariant NMSSM (i.e. with no dimensionful parameters in the superpotential), the $\mu$ parameter is generated dynamically when the singlet obtains a vev: $\mu\equiv \lambda v_s$ (for a discussion of other NMSSM incarnations, such as those with an explicit $\mu$ term, see e.g. Ref.~\cite{Ellwanger:2009dp}).  $\mu$ dictates the Higgsino fraction of $\chi_1^0$, and so $g_{h_1\chi\chi}$ (and $\sigma_{\rm SI}$) are increased for smaller values.  However, $\mu$  cannot be arbitrarily small.  First, LEP restricts $|\mu|\gtrsim 100$ GeV and the LHC further implies $|\mu|\gtrsim 150$ GeV in our scenarios, as discussed in Sec.~\ref{subsec:constraints}.  Secondly, a large Higgsino fraction in the lightest neutralino generally increases the coupling $g_{h_2\chi\chi}$ and correspondingly the invisible branching fraction and total width of $h_2$.  Once again we find tension between obtaining a 126 GeV Higgs in agreement with observation and a large $\sigma_{\rm SI} \gtrsim 10^{-42}$ cm$^2$.  However, there is a way out in this case: for certain values of $\kappa$, an accidental cancellation can occur to reduce the value of $g_{h_2\chi \chi}$.  We find this to be the case if we choose negative values of $\kappa$ in the range $-0.3\lesssim \kappa \lesssim -0.2$.  Choosing $\kappa$ to fall in this range will allow for smaller values of $\mu$, and hence larger values of $\sigma_{\rm SI}$.  

The parameter $\kappa$ also governs the dark matter annihilation rate.  If $\kappa$ is too small, the coupling $g_{h_1h_1h_1}$, which enters into Eq.~(\ref{eq:sigmah}) will not be large enough to effectively drive down the relic neutralino density to the range observed by PLANCK and WMAP.  We typically find that $|\kappa|\gtrsim 0.2$ is required to drive down the relic abundance to the observed range, which encompasses the range of $\kappa$ required for a significant cancellation to occur in $g_{h_2\chi\chi}$ for small values of $\mu$.  The precise choice of $\kappa$ also affects the $h_2 h_1 h_1$ coupling (and hence $\Gamma_{h_2}^{tot}$) -- we discuss this in more detail below.



\begin{figure*}[!t]
\mbox{\includegraphics[width=0.5\textwidth,clip]{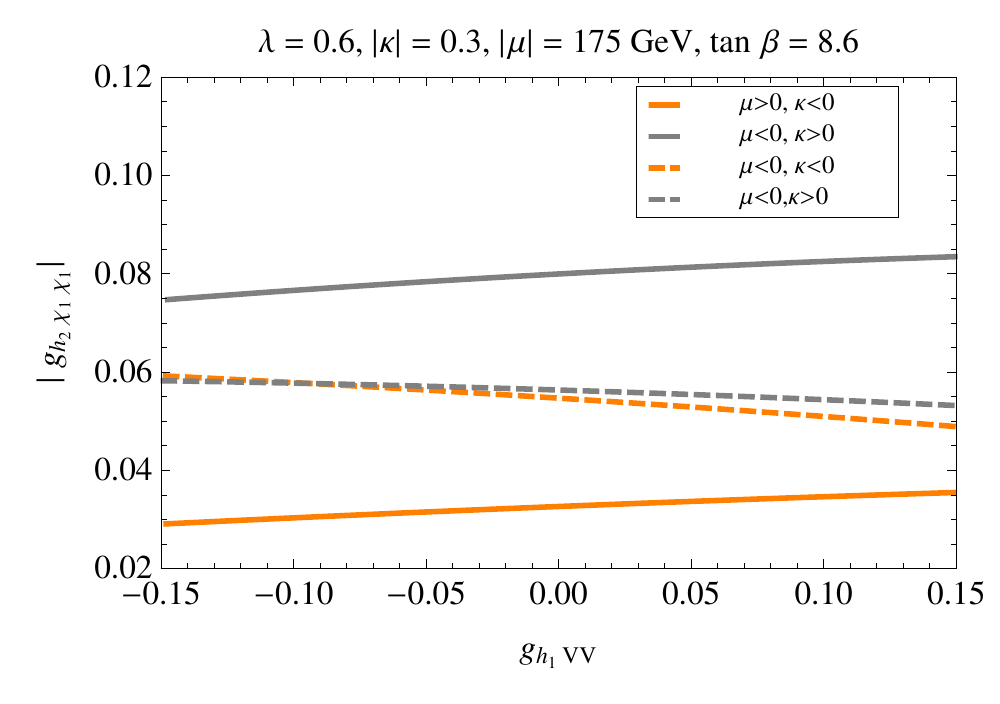}\hspace{-.1cm} \includegraphics[width=0.5\textwidth,clip]{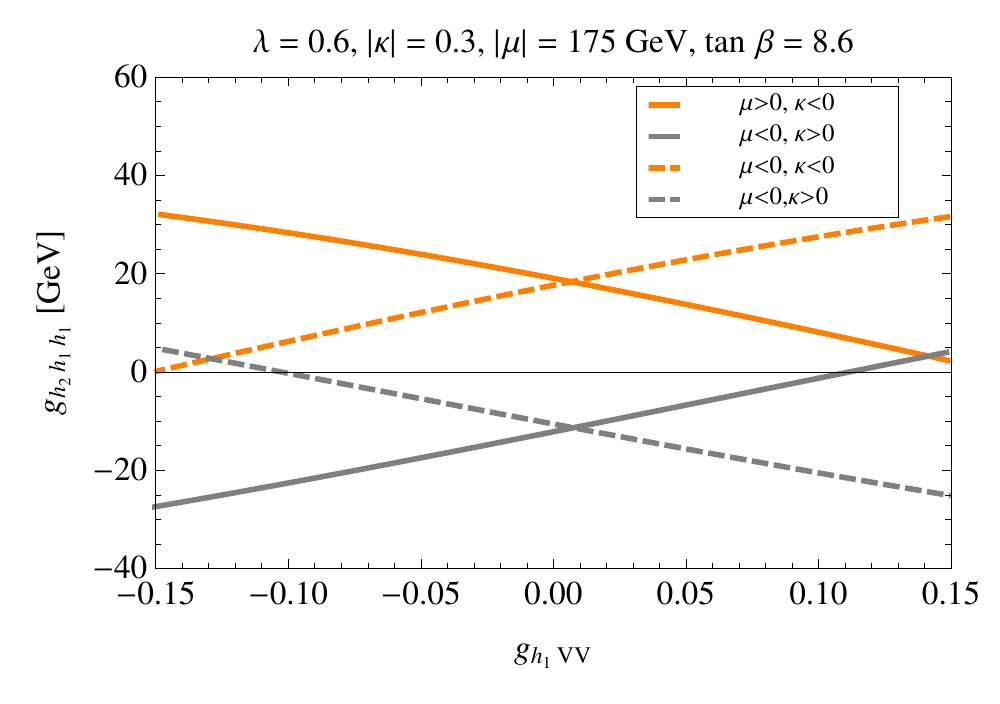}}
\caption{ \label{fig:couplings}\it\small Couplings of $h_2$ to neutralinos (Left) and singlet-like Higgs scalars (Right) as a function of $g_{h_1 VV}$ and the different possible sign choices for $\mu$ and $\kappa$.  The remaining parameters are chosen as in Sec.~\ref{sec:choose} so that $m_{h_1}=6$ GeV, $m_{h_2}=125$ GeV, and $m_{\chi}=10$ GeV with $m_{sf}=2.5$ TeV and $A_i=2500$ GeV.  As discussed in the text, $\left|g_{h_2 \chi \chi}\right|$ is typically smallest for $\mu>0$ and $\kappa<0$ in this scenario. Note that in each case, there is a particular range of $g_{h_1VV}$ for which the $h_2h_1h_1$ coupling becomes small, thus reducing the $h_2\rightarrow h_1 h_1$ partial width. }
\end{figure*}

Taking all of this into account, and requiring $-0.3\lesssim \kappa \lesssim 0.2$, we find that $\left|\mu\right|$ can be in the range $170\leq \left|\mu\right|\leq 220$ GeV and be consistent with the observed Higgs boson and the CDMS II results. We will use $\mu=174$ GeV for our scan in the following section.  Positive values of $\mu$ tend to result in smaller $g_{h_2\chi\chi}$ from partial cancellations between the last two terms in Eq.~(\ref{eq:h2neutcoupling}) with $\kappa<0$, but, with some care, positive values might be chosen just as well.  To illustrate this, we plot $ g_{h_2\chi\chi}$ as a function of the $h_1$ coupling to gauge bosons for various values of $\mu$ on the left side of Fig.~\ref{fig:couplings}.  Here $\lambda=0.59$, $\kappa=-0.297$, $\tan\beta=8.6$ and the other parameters chosen so that $m_{\chi}=10$ GeV, $m_{h_1}=6$ GeV, and $m_{h_2}\approx 125.5$ GeV.  Our procedure for determining the remaining parameter values are described below.

Since $\kappa$ is not too small, the singlino mass, $\sim \kappa \mu/\lambda$ will typically be too large in this case to result in a $\sim 10$ GeV singlino-like lightest neutralino.  LEP constraints dictate $M_2\gtrsim 100$ GeV, so we must have $|M_1| \sim 10$ GeV to obtain a light enough neutralino (which will thus be bino-like).  We take $M_2=650$ GeV so that LHC constraints on associated production of wino-like charginos and neutralinos are satisfied.  $M_1$ can also take on both signs, which will affect the cross-sections relevant for direct detection and neutralino pair annihilation at finite temperature (note, however, that $M_1<0$ tends to increase the invisible branching fraction of $h_2$ given our choices for the other parameters).  To set $M_1$, we diagonalize $\mathcal{M}_{\chi^0}$ (including the leading one-loop corrections) and set the lightest neutralino mass to $m_{\chi_1}=10$ GeV as a mass representative of the CDMS II best-fit region  (to consider $M_1<0$ one must alternatively impose $m_{\chi_1}<0$).  Thus, across the parameter space we consider, the lightest neutralino mass will be fixed.  

For our parameter space study, we would like to vary both the mass of $h_1$ and the coupling of $h_1$ to $SU(2)$ gauge bosons relative to that of the Standard Model Higgs, $g_{h_1 V V}$, since the latter quantity is constrained by LEP searches to be $g_{h_1 V V}^2\lesssim 2\times 10^{-2}$ for $m_{h_1}\approx 6$ GeV \cite{ALEPH, Abbiendi:2002qp}.  This reduced coupling is given by
\begin{equation}
g_{h_1 V V} = S_{11}\cos\beta + S_{12} \sin\beta
\end{equation}
in our conventions.

To see how one might vary $g_{h_1 V V}$, we can use the fact that $h_v^0$ carries the tree-level SM Higgs-gauge boson couplings  and consider for the moment the case of $\tan \beta \sim 1$, whereby the $h_v^0-H_v^0$ and $h_s^0-H_v^0$ mixing is negligible.  Then the problem reduces to the CP-even parameter space specified by the $2\times 2$ $h_v^0-h_s^0$ mixing matrix 
\begin{equation} \label{eq:HS} 
\mathcal{M}_{S,2\times2}^2=\left(\begin{array}{cc}\mathcal{M}_{S,11}^2&\mathcal{M}_{S,13}^2\\ \mathcal{M}_{S,13}^2&\mathcal{M}_{S,33}^2 \end{array} \right). 
\end{equation} 
Here, the matrix entries should include the relevant quantum corrections to the Coleman-Weinberg potential and kinetic terms in the effective action, described in Ref.~\cite{Ellwanger:2009dp}.  

We can diagonalize $\mathcal{M}_{S,2\times2}$ by rotating through by an angle $\theta$ given by 
\begin{equation} \label{eq:theta}
\theta=\frac{1}{2}\tan^{-1}\left(\frac{\mathcal{M}_{S,13}^2}{\mathcal{M}_{S,11}^2-\mathcal{M}_{S,33}^2}\right)
\end{equation}
which yields the physical eigenstates 
\begin{align}
h_1&=h_v^0\sin\theta +h_s^0\cos\theta\\ 
h_2&=h_v^0 \cos\theta -h_s^0\sin\theta 
\end{align}
with corresponding masses
\begin{equation} \label{eq:masses}
m_{h_{1,2}}^2=\frac{1}{2}\left(\mathcal{M}^2_{S,11}+\mathcal{M}^2_{S,33}\pm \sqrt{\left(\mathcal{M}^2_{S,11}-\mathcal{M}^2_{S,33}\right)^2 +4 \mathcal{M}_{S,13}^4}\right),
\end{equation}
provided that the singlet-like state is the lighter of the two.  The angle $\theta$ quantifies the $h_v^0-h_s^0$ mixing, with $\theta=0$ corresponding to unmixed eigenstates.  In the $\tan\beta=1$ limit, $g_{h_1 V V}=\sin\theta$ and the bounds from searches for light Higgs bosons at ALEPH and OPAL dictate that \cite{ALEPH, Abbiendi:2002qp} $\sin^2\theta\lesssim 2\times10^{-2}$ for $m_{h_1}\approx 6$ GeV.

Since we are working with $\tan\beta$ larger than 1, the $h_s^0-H_v^0$ mixing does not entirely vanish and so $\sin\theta$ does not precisely correspond to reduced coupling $g_{h_1 VV}$; one must instead diagonalize the full $3\times3$ system.  However, $g_{h_1 VV}$ still generally varies linearly with $\sin \theta$ and so by considering $\sin\theta$ over an appropriately chosen range, we can scan over all relevant values of $g_{h_1 VV}$.  For example, for the scenario presented in Fig.~\ref{fig:CPE} and in our first benchmark below, scanning over $-0.15\leq g_{h_1VV}\leq 0.15$ amounts to varying $-0.55\lesssim \sin\theta \lesssim 0.55$.  In diagonalizing $\mathcal{M}_S^2$ we use the full 1-loop plus leading two-loop results as implemented in \texttt{NMSSMTools 4.0.0} \cite{NMSSMTools}.  Note that, for the ranges of parameters we consider, the $h^0_v-H^0_v$ mixing is a small effect, and so the $h_2$ reduced couplings $\kappa_Y$ are primarily dictated by $\sin \theta$, with the tree-level effective couplings coming very close to their SM values for small values of $\sin\theta$ (the photon and gluon reduced couplings will depend on the rest of the spectrum; see the discussion below).

Given these simple expressions above, we can now vary $m_{h_1}$ and $\sin \theta$ over the appropriate range and use Eqs.~(\ref{eq:MS}) (including quantum corrections),~(\ref{eq:theta}) and~(\ref{eq:masses}) to solve for the required values of $A_{\lambda}$ and $A_{\kappa}$ (there is typically a unique solution corresponding to a singlet-like $h_1$).  Once $A_{\lambda}$ and $A_{\kappa}$ are determined (along with $\lambda$, $\kappa$, $\tan \beta$, $M_1$, $M_2$, and $\mu$ chosen as described above), the full neutralino and Higgs matrices can be diagonalized to obtain the corresponding diagonalizing matrices $N_{ij}$ and $S_{ij}$, respectively, which enter into the couplings $g_{h_1 h_1 h_1}$, $g_{h_1 \chi_1 \chi_1}$, as well as the effective neutralino-quark couplings $a_{q_1}$.  Then the cross-section $\sigma_{\rm SI}$ and dark matter relic density $\Omega h^2$ can be computed using the expressions found in Sec~\ref{sec:light} and the relevant constraints described in Sec~\ref{subsec:constraints} imposed.  We summarize our choices for the various parameters and the motivation behind them in Table~\ref{tab:choices}

\begin{table}[tc]
\centering

 \begin{tabular}{| c | c | c |}
\hline
Parameter & Approximate Range & Motivation \\
\hline
$\lambda$ &$ \left[0.5,0.6\right]$ & Sizable $\sigma_{\rm SI}$, small invisible $h_2$ branching fraction \\
$\tan\beta$ &$ \left[5, 10\right]$ & Sizable $h_1$ coupling to down-type fermions, $\Upsilon$ decays\\
$\left|\mu\right|$ & $\left[\right.$150 GeV, 300 GeV$\left.\right]$& Chargino and neutralino searches, sizable $\sigma_{\rm SI}$ \\
$\left|\kappa\right|$ & $\left[0.2, 0.5\right]$& Neutralino relic density\\
$\left|M_1\right|$ & $\left[5\right.$ GeV,20 GeV$\left.\right]$& Lightest neutralino mass compatible with CDMS II\\
$M_2$ & 650 GeV & LHC electroweakino searches\\
$A_{\kappa}$& -- & Determined by $m_{h_1}$, $g_{h_1 VV}$\\
$A_{\lambda}$& -- &Determined by $m_{h_1}$, $g_{h_1 VV}$\\
\hline

\end{tabular}
\caption{\label{tab:choices} \it\small The range of parameters considered for the light CP-even scenario, as well as the motivation behind each choice.  The ranges presented are approximate and were determined heuristically by performing several scans with sfermion masses and mixing in the $0.75-3$ TeV range.  The gluino mass, which is unimportant for the dark matter phenomenology, is set to $M_3=3.8$ TeV. }
\end{table}

\subsection{Results}

\begin{figure*}[!t]
\mbox{\includegraphics[width=0.55\textwidth,clip]{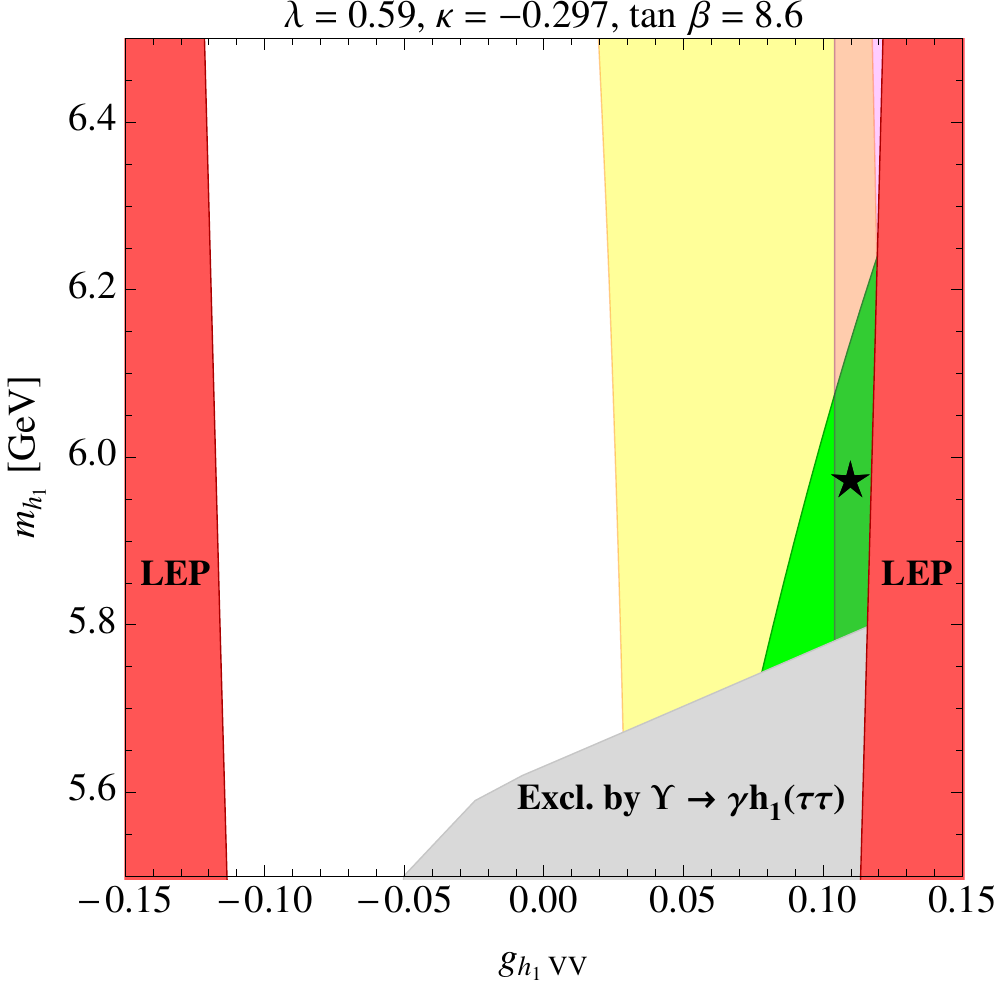}}
\caption{\label{fig:CPE}\it\small Parameter space for the light CP-even case with $\lambda=0.59$, $\kappa=-0.297$, $\tan\beta=8.6$, $m_{\chi}=11$ GeV, $m_{\widetilde{t}}=2.5$ TeV, and $A_{Q,U}=2.5$ TeV, with constraints.  The green region features $1\times 10^{-42}$ cm$^2\leq \sigma_{\rm SI}\leq 2\times 10^{-41}$ cm$^2$, corresponding to the 2$\sigma$ region for the CDMS signal with an $11$ GeV neutralino.  Points in the narrow magenta band have  $\left|g_{h_2h_1h_1}\right|\leq3$ GeV.  The viable region, with $g_{h_2 h_1 h_1}\leq 1$ GeV and the correct relic density, is shown in the darker green.  Gray regions are excluded by radiative $\Upsilon$ decays.  LEP constraints on $h_1$ production exclude the red shaded regions, corresponding to $\left|g_{h_1 VV}\right|\lesssim 0.12$ for $m_{h_1}\approx 6$ GeV. The reduced coupling of $h_2$ to SM gauge bosons falls in the range $\kappa_V\in \left[0.96,1.0\right]$ across the region shown. The benchmark point in Table~\ref{table:bm1} is marked with a star.  The SM-like Higgs mass falls near $m_{h_2}\sim 126$ GeV across the region shown.}
\end{figure*}

Using the strategy outlined above, we can consider a sample portion of the parameter space to illustrate the light CP-even scenario.  We perform a scan as described in the preceding sub-section, varying $m_{h_1}$ and $g_{h_1 VV}$, with $\lambda=0.59$, $\kappa=-0.297$, $\tan\beta=8.6$, $m_{\chi}=11$ GeV, $\mu=174$ GeV, $M_2=650$ GeV, $m_{\widetilde{t}}=2.5$ TeV, $A_{Q,U}=2.5$ TeV and show the results (with the relevant constraints) in Fig.~\ref{fig:CPE}.  These choices yield a SM-like Higgs with mass $m_{h_2}\sim 126$ GeV across the parameter space depicted.  LEP searches for a light $h_1$ exclude the regions shown in red, corresponding to $g_{h_1 VV}\lesssim 0.12$ for $m_{h_1}\approx 6$ GeV.  Meanwhile, the green shaded regions have $\sigma_{\rm SI}$ in the $2\sigma$ best fit CDMSII region for $m_{\chi}\approx 11$ GeV, $1\times 10^{-42}$ cm$^2\leq \sigma_{\rm SI}\leq 2\times 10^{-41}$ cm$^2$ \cite{cdms3events}.  As expected from Eq.~(\ref{eq:a_q}), increasing $m_{h_1}$ suppresses $\sigma_{\rm SI}$ below the levels required to explain the CDMS results \cite{cdms3events}.  Also, as the coupling $\left|g_{h_1 V V}\right|$ is increased, $\sigma_{\rm SI}$ is bolstered by an increased coupling of the $h_1$ mediator to quarks.

The yellow band in Fig.~\ref{fig:CPE} features a relic density in the range of Eq.~(\ref{eq:relden}).  Since $\langle \sigma v\rangle$ is dominated by $p$-wave $s$-channel $h_1$ exchange, the process is not resonant and so not highly sensitive to $m_{h_1}$, provided that $m_{h_1}\leq m_{\chi}$ (for $h_1 h_1$ final states).  The finite-temperature annihilation rate is however sensitive to $g_{h_1 VV}$, since a larger coupling typically increases $g_{h_1 \chi \chi}$. 

An important constraint is that coming from the total width of $h_2$.  In order for $\Gamma_{h_2}^{tot}/\Gamma_{h_{SM}}^{tot}\lesssim 2$ as discussed in Sec.~\ref{subsec:constraints}, both $g_{h_2 \chi \chi}$ and $g_{h_2 h_1 h_1}$ must be relatively small.  The former is made small by our choice of $\mu$ and leads to $BR(h_2\rightarrow {\rm invis}.)\lesssim 36\%$ across all the parameter space shown.  However, $g_{h_2 h_1 h_1}$ is only acceptably small in certain specific regions, illustrated by the magenta band in Fig.~\ref{fig:CPE}, in which $\left|g_{h_2 h_1 h_1}\right|\leq 3$ GeV, a requirement for obtaining an acceptable $\Gamma_{h_2}^{tot}$ for our choices of parameters.  The $h_2 h_1 h_1$ coupling in this region is minimized by cancellations between different terms in the expression for $g_{h_2 h_1 h_1}$ \cite{Ellwanger:2009dp}:
\begin{equation}\label{eq:h2h1h1}
\begin{aligned}
g_{h_2h_1h_1}=&\frac{\lambda^2}{\sqrt{2}}\left[ v\cos\beta \left(\pi^{122}+\pi^{133}\right)+v\sin\beta\left(\pi^{211}+\pi^{233}\right)+\frac{\mu}{\lambda}\left(\pi^{311}+\pi^{322}\right)\right]\\
&-\frac{\lambda \kappa}{\sqrt{2}}\left(v\cos\beta \pi^{323}+v\sin\beta \pi^{313}+2\frac{\mu}{\lambda}\pi^{123}\right)+\frac{\sqrt{2}\kappa^2 \mu}{\lambda}\pi^{333}-\frac{\lambda A_{\lambda}}{\sqrt{2}}\pi^{123} +\frac{\kappa A_{\kappa}}{3\sqrt{2}}\pi^{333}\\
&+\frac{g^2}{2\sqrt{2}}\left[v\cos\beta \left(\pi^{111}-\pi^{122}\right)-v\sin\beta \left(\pi^{211}-\pi^{222}\right)\right]
\end{aligned}
\end{equation}
where $\pi^{ijk}$ is the sum of all six permutations of the indices $a,b,c$ in $S_{ai}S_{bj}S_{ck}$, where $a=2$, $b=c=1$.  The largest contributions to $g_{h_2h_1h_1}$ in our case come from the terms proportional to $\lambda^2$ and $\lambda \kappa$ in Eq.~(\ref{eq:h2h1h1}).  Significant cancellations can occur between these terms for certain values of $\kappa$.  This can be seen on the right hand side of Fig.~\ref{fig:couplings}, where we plot $g_{h_2h_1h_1}$ as a function of $g_{h_1 VV}$ for various sign choices of $\mu$ and $\kappa$, with other parameters chosen such that $m_{\chi}=10$ GeV, $m_{h_1}=6$ GeV, and $m_{h_2}=125.5$ GeV with $m_{\widetilde{t}}=2.5$ TeV and $A_{Q,U}=2.5$ TeV.  The coupling of $h_2$ to the singlet-like Higgs can become very small, thereby reducing the branching fraction of $h_2$ to $h_1h_1$ pairs and the total width of $h_2$.  As noted previously, the relatively small value of $g_{h_2h_1h_1}$ also allows the LEP and Tevatron constraints on $h_1$ production from $h_2$ decays to be satisfied by a significant margin.  Without these cancellations, $\Gamma_{h_2}^{tot}$ is generally unacceptably large, so this requirement is quite crucial for the light CP-even scenario.  As we will see in the following Section, the light CP-even/CP-odd case typically requires both $\lambda$ and $\kappa$ to be smaller, and so such cancellations are not typically required.

The gray shaded regions in Fig.~\ref{fig:CPE} are excluded by radiative $\Upsilon$ decays to taus (this is the most constraining channel).  The excluded regions correspond to sizable positive $g_{h_1 VV}$ since here the $h_1$ coupling to down-type fermions also tends to be larger.  All other flavor physics, collider, and Higgs constraints are satisfied across the parameter space.  $B$ physics constraints are satisfied, since on-shell decays of $B$ through $h_1$ are prohibited by virtue of considering $m_{h_1}\gtrsim 5$ GeV.  Meanwhile, the reduced coupling of $h_2$ to SM gauge bosons, $\kappa_V$, remains in the interval $\left[ 0.98,1\right]$ across the entire parameter space and all the reduced couplings fall within the 95\% C.L. regions outlined in Ref.~\cite{Belanger:2013xza}.

To show more clearly what this scenario entails, we present the details of a benchmark point from the scan in Table~\ref{table:bm1} (the benchmark point is indicated with a star in Fig.~\ref{fig:CPE}).  We have checked that this point satisfies all collider constraints in \texttt{NMSSMTools 4.0} and \texttt{HiggsBounds 4.0} through a call to \texttt{micrOmegas 3.2} as well as all flavor physics constraints implemented in \texttt{NMSSMTools}.  The various reduced couplings of $h_2$ and $\Gamma_{h_2}^{tot}$, $BR(h_2\rightarrow {\rm invis.})$ for this point fall within the 95$\%$ C.L. regions of the global Higgs fit in Ref.~\cite{Belanger:2013xza}.  The largest deviation from the best fit reduced couplings in this case is in the loop-induced couplings of $h_2$ to photons, with $\kappa_{\gamma}\sim 1$. Further modifications of the rest of the sfermion spectrum (by e.g. including light staus \cite{light_staus}) may be used to raise the reduced coupling while leaving the rest of the phenomenology in tact.  This may be required if the increased $h_2\rightarrow \gamma \gamma$ signal strength relative to the SM \cite{Atlas_couplings} persists in the ATLAS data (however with recent CMS results \cite{CMS_couplings} this is looking less likely).  We leave this possibility to future study.  We emphasize, however, that the `low' diphoton rate is not an inherent feature of this scenario.  Note also that the spin-dependent scattering cross-section in this case is rather large but experimentally allowed (see e.g. Ref.~\cite{Buckley:2013gjo}).  Also, LHC Mono-jet search constraints are satisfied by a significant margin, checking against the results of e.g. Ref.~\cite{Cotta:2013jna} (the decay $h_1\rightarrow \chi_1 \chi_1$ is kinematically forbidden). 


\begin{table}[tc]
\centering

 \begin{tabular}{|| c || c || c || c || c ||}
\hline \hline
  $\lambda$  &   $\kappa$   & \hspace{.3 cm}  $\tan \beta$ \hspace{.3 cm} &\hspace{.05 cm}  $A_{\lambda}$ [GeV] \hspace{.05 cm} &  \hspace{.05 cm} $A_{\kappa}$ [GeV]  \hspace{.05 cm}\\  \hline

0.59 & -0.297 & 8.6 & 1867.9 & 404.72 \\
\hline\hline
$\mu$ [GeV]  \hspace{.05 cm} & \hspace{.05 cm}  $M_1$ [GeV] \hspace{.05 cm}  & \hspace{.05 cm}  $M_2$ [GeV] \hspace{.05 cm}&  \hspace{.05 cm}  $m_{h_1}$ [GeV] \hspace{.05 cm}  & \hspace{.05 cm}  $m_{h_2}$ [GeV] \hspace{.05 cm}    \\
\hline
174.0& 10.8 & 650 &6.0& 123.8 \\
\hline \hline
\hspace{.05 cm}  $m_{a_1}$ [GeV] \hspace{.05 cm}  &\hspace{.05 cm}  $m_{a_2}$ [GeV] \hspace{.05 cm}  & \hspace{.05 cm}  $m_{\chi_1}$ [GeV] \hspace{.05 cm}  & $\Omega h^2$ & $\sigma_{\rm SI}$ [cm$^2$]\\
\hline
  316.1 & 1610.0 & 11.0 & 0.123 & $2.6 \times 10^{-42}$  \\
   \hline \hline
$\sigma ^{\rm SD}$ [cm$^2$]&$\sigma v$ [cm$^3/$s] &  $\Gamma_{h_2}^{tot}/\Gamma_{h_{SM}}^{tot}$ & $BR(h_2\rightarrow h_1 h_1)$ & $BR(h_2 \rightarrow \chi_1^0 \chi_1^0)$\\ \hline
 $8.4 \times 10^{-41}$& $2.2 \times 10^{-29}$ & 1.3 & 2.9\% & 34.3\% \\
\hline \hline



\end{tabular}
\caption{\label{table:bm1} \it\small Benchmark point for a light bino-like neutralino with $\sigma_{\rm SI}$ in the current $2 \sigma$ best-fit region for the CDMS II results.  The stop masses are all set to 2.5 TeV, the other squarks are at 2 Tev, the sleptons have mass 1.8 TeV, and $A_{Q,U}=2.5$ TeV, $A_l=1$ TeV (we use the same slepton prameters throughout this work).  The gluino mass, which is unimportant for the dark matter phenomenology, is set to $M_3=3.8$ TeV.  All couplings of $h_2$, as well as its invisible branching fraction and total width, fall within the $95\%$ C.L. regions suggested by the global fit in Ref.~\cite{Belanger:2013xza}.}
\end{table}

The benchmark in Table~\ref{table:bm1} shows that the light CP-even scenario is on the verge of being in significant tension with constraints on the invisible branching fraction and total width of $h_2$.  We have performed scans for several other choices of parameters in the range outlined in Table~\ref{tab:choices} and have found it generally quite difficult to reduce $\Gamma_{h_2}^{tot}/\Gamma_{h_{SM}}^{tot}\lesssim 1.4$ and $BR(h_2\rightarrow {\rm invis}.)\lesssim 30\%$ while achieving a large spin-independent neutralino-nucleon scattering cross-section.  

In reaching the above conclusions, we have demanded $m_{h_1}\gtrsim 5$ GeV.  However, mediator masses close to the charmed resonances might still be allowed observationally, since the $B$ meson experiments veto in these mass ranges \cite{LHCb, BELLE, BaBar}.  In this case, on-shell $B$ decays to muons will be allowed, but effectively hidden.  Still, one has to contend with $\Upsilon$ decay constraints, and the limits are more stringent for smaller masses.  Since the couplings of $h_1$ to gauge bosons (and up-type fermions) must be small, the coupling to down-type fermions cannot be too small if we still hope to achieve $\sigma_{\rm SI} \gtrsim 10^{-42}$ cm$^2$.  However, inspecting the constraints from Ref.~\cite{ups_hadronic} (the hadronic final states are the most constraining below the $\tau$ threshold), there are several bins of $m_{h_1}$ for which larger couplings of $h_1$ to down-type fermions are technically allowed, particularly near the $J/\psi$ resonance, with $m_{h_1}\simeq 3.1$ GeV \cite{ups_hadronic}.  Since the $m_{h_1}$ is smaller in this case, $\sigma_{\rm SI}$ can be larger than for $m_{h_1}\gtrsim 5$ GeV.  We illustrate this possibility with another benchmark in Table~\ref{table:bm2}.  We see that, since $\lambda$ can be smaller in this case, the properties of $h_2$ can be in better agreement with the resonance observed at the LHC, with a smaller invisible branching fraction and $BR(h_2\rightarrow h_1 h_1)$.  We emphasize, however, that this is a very highly tuned scenario, requiring a conspiracy of parameters to allow $m_{h_1}$ to fall in the narrow range allowed by $B$ and $\Upsilon$ meson experiments.  Still, it is in principle possible for such a light mediator to have escaped detection thus far and provide larger spin-independent neutralino-nucleon scattering cross-sections.

\begin{table}[tc]
\centering

 \begin{tabular}{|| c || c || c || c || c ||}
\hline \hline
  $\lambda$  &   $\kappa$   & \hspace{.3 cm}  $\tan \beta$ \hspace{.3 cm} &\hspace{.05 cm}  $A_{\lambda}$ [GeV] \hspace{.05 cm} &  \hspace{.05 cm} $A_{\kappa}$ [GeV]  \hspace{.05 cm}\\  \hline

0.5 & -0.23 & 8.6 & 1832.0 & 363.03 \\
\hline\hline
$\mu$ [GeV]  \hspace{.05 cm} & \hspace{.05 cm}  $M_1$ [GeV] \hspace{.05 cm}  & \hspace{.05 cm}  $M_2$ [GeV] \hspace{.05 cm}&  \hspace{.05 cm}  $m_{h_1}$ [GeV] \hspace{.05 cm}  & \hspace{.05 cm}  $m_{h_2}$ [GeV] \hspace{.05 cm}    \\
\hline
175.2 & 8.2 & 650 &3.1 & 124.2 \\
\hline \hline
\hspace{.05 cm}  $m_{a_1}$ [GeV] \hspace{.05 cm}  &\hspace{.05 cm}  $m_{a_2}$ [GeV] \hspace{.05 cm}  & \hspace{.05 cm}  $m_{\chi_1}$ [GeV] \hspace{.05 cm}  & $\Omega h^2$ & $\sigma_{\rm SI}$ [cm$^2$]\\
\hline
   289.44 &1600.86 & 7.9 & 0.131 & $1.2 \times 10^{-41}$  \\
   \hline \hline
$\sigma ^{\rm SD}$ [cm$^2$]&$\sigma v$ [cm$^3/$s] &  $\Gamma_{h_2}^{tot}/\Gamma_{h_{SM}}^{tot}$ & $BR(h_2\rightarrow h_1 h_1)$ & $BR(h_2 \rightarrow \chi_1^0 \chi_1^0)$\\ \hline
 $8.4 \times 10^{-41}$& $2.1 \times 10^{-29}$ & 1.3 & 9.4\% & 14.6\% \\
\hline \hline

\end{tabular}
\caption{\label{table:bm2} \it\small Benchmark point for a light bino-like neutralino with $\sigma_{\rm SI}$ in the current $1\sigma$ best-fit region for the CDMS II results. This point features a light scalar hidden under the $J/\psi$ resonance   The stop masses are set to $m_{sq}=2.5$ TeV, the other squarks to 2 TeV, with triscalar couplings $A_{t,b}=2.5$ TeV (the slepton parameters and gluino mass are as before).  Although this point features a rather large coupling of $h_1$ to down-type fermions ($S_{11}/\cos\beta\approx 0.5$), there is a significant, though very narrow, upward fluctuation in the $\Upsilon$ decay limits at the $J/\psi$ resonance, which allows this point to lie just below the exclusion limits.  On-shell $B$ decays through $h_1$ into muons are not constraining, since all experiments veto dimuon invariant masses near the $J/\psi$ resonance. All couplings of $h_2$, as well as its invisible branching fraction and total width, fall within the $95\%$ C.L. regions suggested by the global fit in Ref.~\cite{Belanger:2013xza} for the most conservative case of all Higgs couplings as in the SM but allowing for invisible decays.  This point satisfies all constraints from \texttt{HiggsBounds}, as well as all flavor physics constraints as implemented in \texttt{NMSSMTools}.}
\end{table}

Clearly, modest improvements in $\Upsilon$ decay measurements should be able to access the remaining available parameter in the cases we have considered.  Additionally, ongoing LHC efforts at high luminosity will continue to probe the light CP-even scenario quite effectively, both through increased sensitivity to invisible decays and indirect bounds on the total Higgs width.  This can be appreciated by the following crude argument: it is not unreasonable for the high luminosity LHC to infer the Higgs reduced couplings $\kappa_i$ to within $\sim 5-10\%$ precision \cite{slides}.  For the cases we considered, if the Higgs resonance remains consistent with the SM prediction to this point, the observed signal strength for all channels $X\bar{X}\rightarrow h_2 \rightarrow Y\bar{Y}$, must satisfy $\mu_{XY}\lesssim 0.87-1.05$, provided $\Gamma_{h_2}^{tot}/\Gamma_{h_{SM}}^{tot}\gtrsim 1.4$ (as we found for the points we considered).  This is already in tension with current limits on e.g. the diphoton signal strength and so would likely be either confirmed or excluded by the high luminosity LHC.  We defer a more detailed study of the LHC reach for this scenario to future work, but emphasize that as the upper limits on the Higgs invisible branching fraction and total width become more stringent, one will likely be forced to consider smaller values of $\lambda$ in this scenario which will tend to reduce $\sigma_{\rm SI}$.  However, at this point, light NMSSM neutralinos with a light CP-even Higgs remain a viable explanation for the CDMS II events.

 \section{The Light CP-even/CP-odd Scenario: Annihilation through a Light Pseudoscalar} \label{sec:pseudo}

Let us now turn to the light CP-even/CP-odd case.  At zero temperature, the dominant contribution to the neutralino pair annihilation rate is due to the $s$-channel exchange of a pseudoscalar which couples to $b\bar{b}$, given by:
\begin{equation}
 \sigma v_{\chi \chi\rightarrow b\bar{b}} =\frac{N_c\sqrt{s} \left|g_{a_1bb}\right|^2 \left|g_{a_1\chi \chi}\right|^2  \sqrt{s-4m_b^2}}{16 \pi\left(\left(s-m_{a_1}^2\right)^2+\Gamma_{a_1}^2m_{a_1}^2\right)}.
\end{equation}
where the couplings of $a_1$ to $b\bar{b}$ and neutralinos are
\begin{equation}\begin{aligned}
g_{a_1 bb}&=i\frac{m_b}{\sqrt{2}v\cos\beta}P_{11}\\
g_{a_1\chi\chi}&\simeq i \left(\frac{2\lambda}{\sqrt{2}}\left(P_{11} N_{14}N_{15}+P_{12}N_{13}N_{15}+P_{13}N_{13}N_{14}\right)-\sqrt{2}P_{13}N_{15}^2\right)
\end{aligned}
\end{equation}
for a singlet-like $a_1$ and singlino-like $\chi_1^0$.  The width $\Gamma_{A_1}$ is dominated by $A_1\rightarrow b\bar{b}, \chi_1^0\chi_1^0$  and is of order $\sim 10^{-5}$ GeV for points in our scan below.
This $s$-wave annihilation process is resonant at $T=0$ for $2m_{\chi}=m_{A_1}$.  At finite temperature, the thermally averaged pair annihilation rate is given in terms of $ \sigma v$ by Eq.~(\ref{eq:thermal}). Note once again that the zero-temperature resonance gets smeared out at finite temperature.

If the light pseudoscalar is accompanied by a light CP-even Higgs boson, light neutralinos can efficiently annihilate in the early universe and provide a large enough $\sigma_{\rm SI}$ to explain the CDMS II events as before.  Let us once again see what this entails for the parameter space.

\subsection{The Parameter Space}

We impose the following requirements on the spectrum:

\begin{itemize}
\item As before, a Standard Model-like $h_2$ consistent with the resonance observed at the LHC with mass $m_{h_2}\sim 126$ GeV.

\item A lightest neutralino LSP with mass $m_{\chi_1}\sim 5-15$ GeV and with a thermal relic abundance in the range dictated by WMAP and PLANCK, $0.09\lesssim \Omega h^2 \lesssim 0.14$.  We achieve this by requiring a light pseudoscalar near the zero-temperature resonance: $m_{a_1}\approx 2 m_{\chi}$.

\item A large spin-independent neutralino-nucleon elastic scattering cross-section $10^{-42}$ cm$^2 \lesssim \sigma_{\rm SI} \lesssim  10^{-40}$ cm$^2$ as required to explain the CDMS signal \cite{cdms3events}.  To achieve this, we again  require a singlet-like lightest CP-even Higgs with mass $m_{h_1}$ in the range $4.8$ GeV $\lesssim m_{h_1} \lesssim m_{\chi_1}$ and consistent with constraints from the LHC, Tevatron, LEP, and flavor physics.

\end{itemize}

To scan over the parameter space, we will treat $m_{h_1}$ and $m_{a_1}$ as free parameters.  Given values for $\lambda$, $\kappa$, $\tan\beta$, $M_1$, $M_2$ and the sfermion masses and mixing, $A_{\lambda}$, $A_{\kappa}$ can be determined by diagonalizing $\mathcal{M}_S^2$, $\mathcal{M}_P^2$, setting the lowest eigenvalues equal to $m_{h_1}^2$, $m_{a_1}^2$, respectively, and solving the resulting system of equations for $A_{\lambda}$ and $A_{\kappa}$.  This procedure yields unique solutions for $A_{\lambda}$, $A_{\kappa}$ across the parameter space we consider.

The task at hand is thus to determine the remaining parameters suitable for the light CP-even/CP-odd scenario.  As we will see, in contrast to the light CP-even case, viable examples of this scenario typically do not require accidental cancellations or significant tuning in the various couplings to achieve agreement with the constraints outlined in Sec.~\ref{subsec:constraints} and so there is superficially more freedom in choosing parameters.  However, simply requiring a light $h_1,a_1$ and $m_{h_2}=126$ GeV points us to specific regions of the NMSSM parameter space. 

There are two limits of the NMSSM in which a light pseudoscalar appears as a pseudo-Goldstone boson, corresponding to a spontaneously broken symmetry.  In the limit that $\kappa$ vanishes, the superpotential exhibits a $U(1)$ Peccei-Quinn (PQ) symmetry, with the fields transforming as 
\begin{equation}
H_{u,d}^0\rightarrow e^{i\varphi_{PQ}}H_{u,d}^0, \qquad S\rightarrow e^{-2i\varphi_{PQ}}S.
\end{equation}
Alternatively, in the limit of small $A_{\lambda}$, $A_{\kappa}$, the superpotential exhibits a $U(1)$ $R$-symmetry, whereby the fields transform as
\begin{equation}
H_{u,d}^0\rightarrow e^{i\varphi_{R}}H_{u,d}^0, \qquad S\rightarrow e^{i\varphi_{R}}S
\end{equation}
%
%
It is also possible for the pseudoscalar mass to nearly vanish due to accidental cancellations in the CP-odd mass matrix.  This can be seen by taking the determinant of Eq.~(\ref{eq:MA}).  
\begin{figure*}[!t]
\mbox{\includegraphics[width=0.45\textwidth,clip]{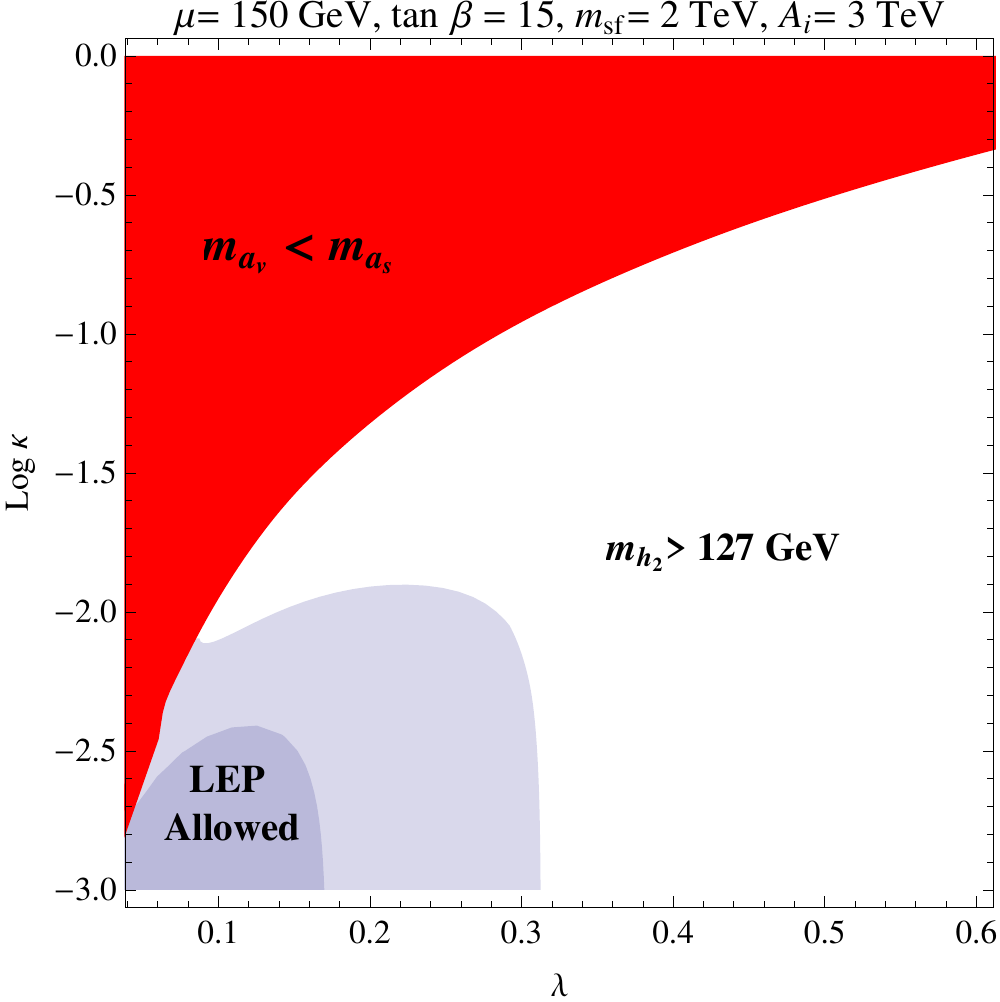}\qquad \includegraphics[width=0.45\textwidth,clip]{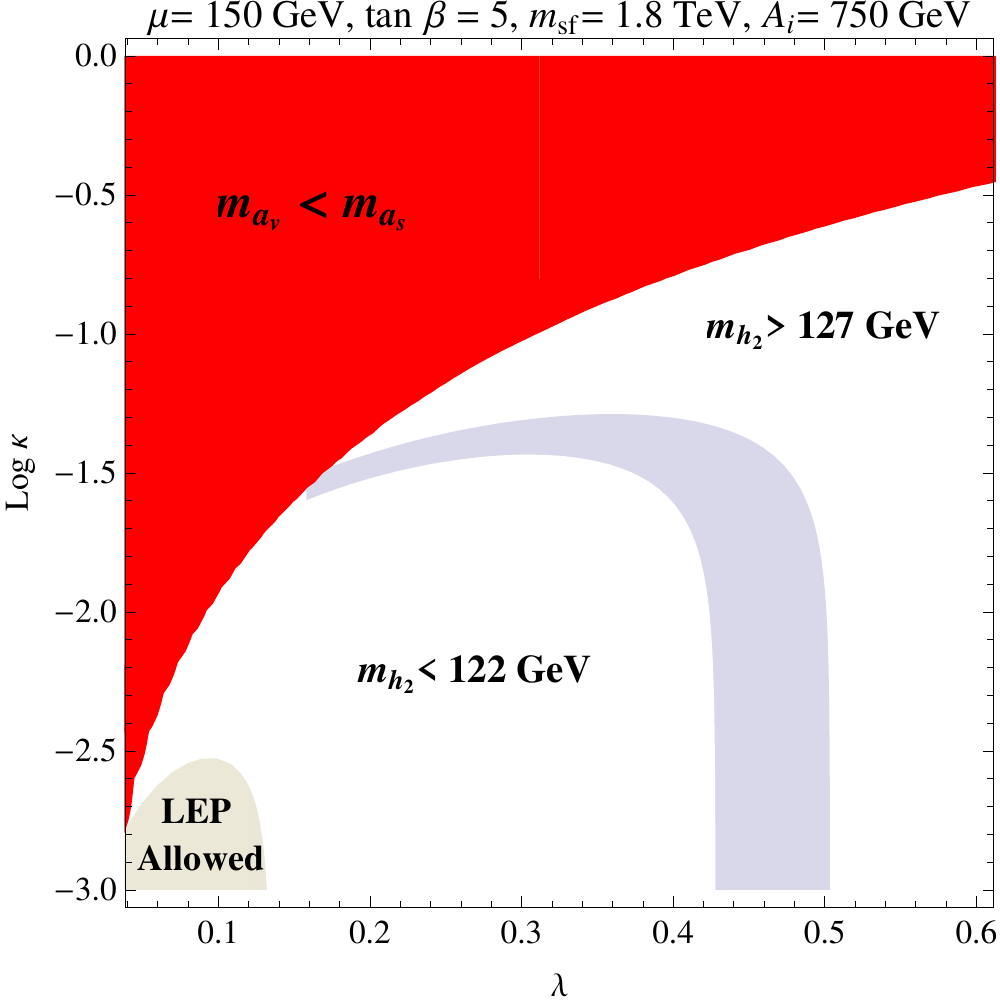}  }
\caption{\label{fig:lk}\it\small Parameter space for the light CP-even/CP-odd scenario for two sets of values of the sfermion masses/mixing, and $\tan\beta$.  The LHS features larger $\tan\beta$ and stops closer to the maximal mixing scenario (see Eq.~(\ref{eq:mix})), while the RHS features smaller $\tan \beta$ and smaller stop mixing.  The red region features a light MSSM-like $a_1$ and is difficult to reconcile with LEP and Tevatron constraints on $h_2$ decays into light $a_1$ pairs.  The light blue shaded regions feature $h_2$ in the range $122$ GeV $\leq m_{h_2}\leq127$ GeV. The dark shaded regions feature light scalars with couplings to gauge bosons allowed by LEP (the white regions are therefore excluded).  A 126 GeV Higgs appears together with a LEP-allowed $h_1$ only in the lower left hand corner of the heavy sfermion case.  The most promising region for the light CP-even/CP-odd scenario is the small-$\lambda$, near-PQ symmetry limit with moderate to large $\tan \beta$.}
\end{figure*}

Additionally requiring a light CP-even Higgs and a 126 GeV SM-like $h_2$ narrows down these possibilities, since a light pseudoscalar and light scalar satisfying current collider constraints appear together only in specific regions of the NMSSM parameter space.  In fact, it is most generic in the small-$\lambda$ regime near the Peccei-Quinn symmetry limit with small values of $\kappa$, $A_{\kappa}$ (this was dubbed the ``Dark Light Higgs" scenario in Ref.~\cite{DLH}). To see this, we show two scans performed over the $\lambda-\kappa$ planes on the right- and left-hand sides of Fig.~\ref{fig:lk} for two representative choices of the sfermion masses, mixing, and $\tan \beta$.  Here we have fixed $m_{h_1}=1$ GeV and $m_{a_1}=20$ GeV (near the $m_{a_1}\approx 2 m_{\chi}$ resonance) for illustration; more realistic choices for these masses yield qualitatively similar results.  The red regions in Fig.~\ref{fig:lk} are those where $\mathcal{M}_{P,11}^2<\mathcal{M}_{P,22}^2$, resulting in $a_1$ being MSSM-like and hence ruled out by e.g. searches for  $h_2\rightarrow a_1 a_1 \rightarrow 4b, 4\tau, 2b 2\tau$ at LEP \cite{Schael:2006cr, Schael:2010aw} and $h_2\rightarrow a_1 a_1 \rightarrow 2\mu^{+}2\mu^-$ decays at the Tevatron \cite{Abazov:2009yi}.   Regions with light scalars whose couplings to the SM gauge bosons is small enough to be consistent with LEP searches are shaded and indicated.  The regions producing a SM-like $h_2$ in the acceptable mass range $122$ GeV $\leq m_{h_2}\leq127$ GeV are shaded light blue.  The only region with an acceptable $h_2$ mass and with $g_{h_1VV}$ allowed by LEP in both scans is found in the larger $\tan\beta$ case on the LHS of Fig.~\ref{fig:lk} in the lower left corner, corresponding to the small-$\lambda$, small-$\kappa$ region.  This region is favored because a small value of $\kappa$ (and $A_{\kappa}$) guarantees a light singlet-like state, while small $\lambda$ tends to reduce the upper bound on the mass of the lightest scalar.  It may be possible to arrange the parameters in such a way that the light CP-even/CP-odd scenario is viable beyond this region, however we focus on the small-$\lambda$, near PQ-symmetric limit for the remainder of our analysis.  Since $\lambda$ is small, the NMSSM tree-level contribution to $m_{h_2}$ is reduced and so larger values of $\tan\beta$ are typically required to bolster the Higgs mass as in the MSSM.  We refer the Reader to Ref.~\cite{DLH} for more details about the phenomenology of this scenario. 

The lightest neutralino is singlino-like in this case, with both $m_{\chi_1}$ and $\sigma_{\rm SI}$ set primarily by the value of $\kappa$: larger values raise $\sigma_{\rm SI}$, but also raise $m_{\chi_1}$.  To lower $m_{\chi_1}$, one might increase $\lambda$ or decrease $\mu$, however both options tend to increase $g_{h_1VV}$.  We thus find tension between the LEP constraints on $h_1$ and obtaining a light neutralino able to explain the CDMS II results.  Heuristically, it appears difficult to lower $m_{\chi_1}$ below $\sim 11$ GeV without producing too small a $\sigma_{\rm SI}$.  From the standpoint of explaining the CDMS II events this is fine, since the best-fit region extends up to $m_{\chi_1}\approx$ 15 GeV, however we note that such `heavy' neutralinos will likely have a difficult time simultaneously explaining the DAMA/LIBRA, CoGeNT, and CRESST-II anomalies if the CDMS II results are ignored.  The Reader should bear this in mind in interpreting our results below.

\subsection{Results}

\begin{figure*}[!t]
\mbox{\includegraphics[width=0.55\textwidth,clip]{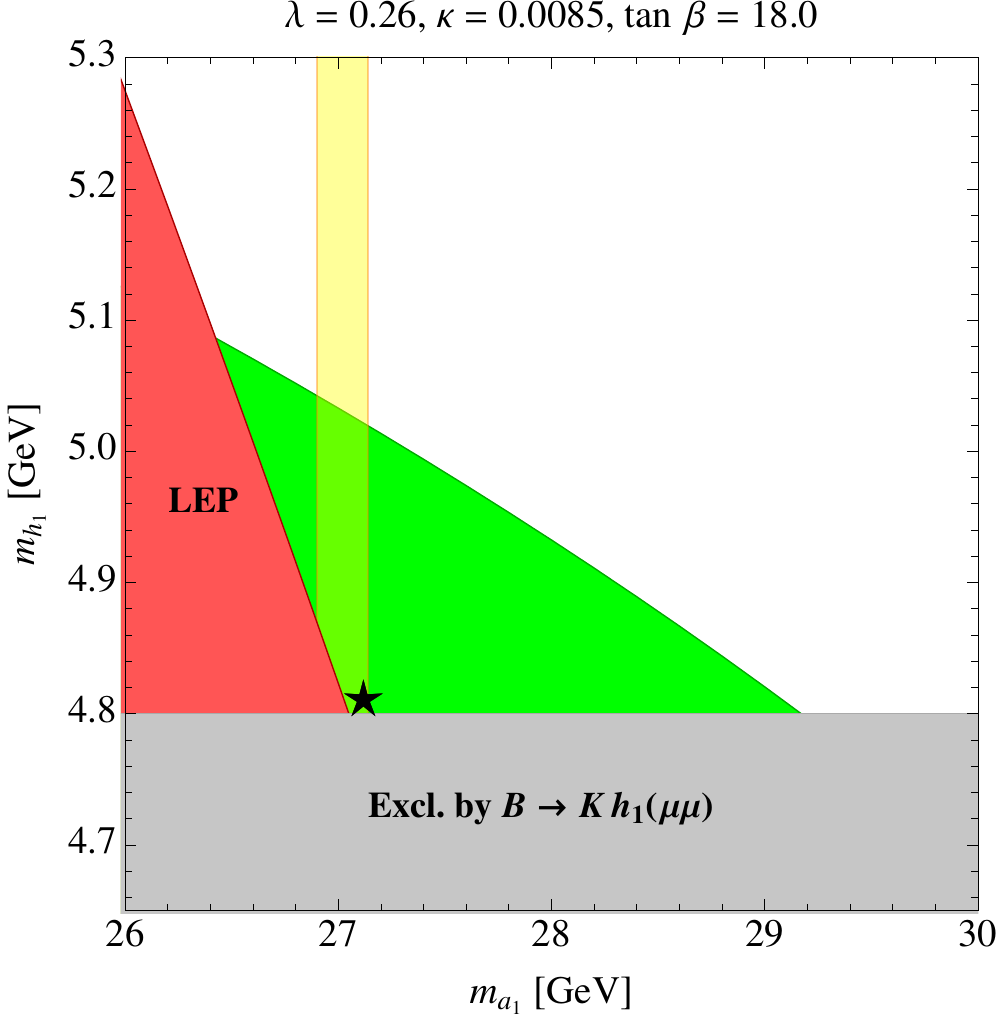}}
\caption{\label{fig:CPO}\it\small Parameter space for the light CP-even/CP-odd case with $\lambda=0.26$, $\kappa=0.0085$, $\tan\beta=18$, $m_{\chi}=11.5$ GeV, $m_{\widetilde{t}}=1$ TeV, and $A_{Q,U}=1.8$ TeV, with constraints.  The green region features $1\times 10^{-42}$ cm$^2\leq \sigma_{\rm SI}\leq 1.5\times 10^{-41}$ cm$^2$, corresponding to the 2$\sigma$ region for the CDMS signal with a $\sim11.5$ GeV neutralino.  The yellow bands contains points with a relic density compatible with PLANCK and WMAP.  The gray region is excluded by on-shell $B\rightarrow K h_1$ decays, with $h_1\rightarrow \mu^{+}\mu^-$, while the red region is excluded by LEP.  Constraints from $\Upsilon$ decays and the $h_2$ couplings/partial widths are satisfied across the parameter space. The reduced coupling of $h_2$ to $SU(2)$ gauge bosons falls in the range $0.99\leq \kappa_{V}\leq 1.00$ for all points shown.  The benchmark point in Table~\ref{table:bm3} is marked with a star.  The SM-like Higgs mass falls near $m_{h_2}\sim 126$ GeV across the region shown.}
\end{figure*}

Focusing on the region outlined above, we show the results of a scan over the $m_{h_1}-m_{a_1}$ plane for representative values of the relevant parameters in Fig.~\ref{fig:CPO}.  We take $\lambda=0.26$, $\kappa=0.0085$, $\tan\beta=18$, $\mu=159$ GeV, $M_1=-139$ GeV, $M_2=800$ GeV, $M_3=1.5$ TeV, $m_{\widetilde{t}}=1$ TeV, $A_{Q,U}=1.8$ TeV and the slepton parameters are as before.  These choices result in $m_{\chi}=11.5$ GeV across the parameter space. The SM-like Higgs mass falls near $m_{h_2}\sim 126$ GeV throughout the region shown.

As in the light CP-even scenario, $\sigma_{\rm SI}$ can fall in the range suggested by CDMS II provided $m_{h_1}$ is small enough.  This can be seen by considering the green region in Fig.~\ref{fig:CPO} which shows points with $1\times 10^{-42}$ cm$^2\leq \sigma_{\rm SI}\leq 1\times 10^{-41}$ cm$^2$, the $2-\sigma$ best-fit region for the scattering cross-section for $m_{\chi}\approx 11.5$ GeV.  Larger values of $m_{h_1}$ again suppress the cross-section.

There are in fact two regions in which the relic density can fall in the appropriate range (although only one can be seen in Fig.~\ref{fig:CPO}).  The first is a small sliver very close to the $m_{a_1}=2m_{\chi}$ resonance (this band falls in the region excluded by LEP).  Here, the thermal average in Eq.~(\ref{eq:thermal}) just begins to pick up the contribution from the resonance, however these points are also solidly excluded by both LEP and dark matter indirect detection results \cite{Ackermann:2012qk, Ando:2012vu}.

The other region resulting in the correct thermal relic abundance is away from the resonance, where $\langle \sigma v\rangle_{T\neq0}$ begins to decrease from the large values found near $m_{a_1}=2m_{\chi}$.  This occurs in the yellow band on the left hand side of the plot in Fig.~\ref{fig:CPO}.  In this region, the zero temperature annihilation rate is small ($\sigma v\sim 10^{-31}-10^{-29}$ cm$^3/$s) and thus not constrained by current dark matter indirect detection experiments.  This also means, however, that the annihilation rate in this scenario is typically too small to account for the excess of gamma-rays coming from the galactic center \cite{Hooper:2012ft, Hooper:2011ti}.  Moving below the resonance, $m_{a_1}<2m_{\chi}$, one might imagine obtaining a large zero-temperature annihilation rate while utilizing $h_1$ to mediate the annihilation in the early universe as in the light CP-even case, however, since $\lambda$ and $\kappa$ are both small, this is not typically possible (see again the discussion in Sec.~\ref{sec:choose}).  Thus, \emph{neither the light CP-even nor CP-even/CP-odd scenarios predict large zero-temperature annihilation rates}.  There are caveats to this statement, however.  For example, the correct relic density may be obtained by some non-thermal production mechanism despite a large pair-annihilation rate \cite{Arcadi:2011ev}.  If a dark matter interpretation of the signal from the Galactic Center is confirmed, one may be led to consider such modifications to the light CP-even/CP-odd case as an explanation for the CDMS II signal.

LEP limits on $h_1$ production exclude the red region in the lower left corner, and on-shell $B\rightarrow K \mu^+\mu^-$ decays exclude the gray region.  All other constraints we consider (outlined in Sec.~\ref{subsec:constraints}) are satisfied across the parameter space shown in Fig.~\ref{fig:CPO}.  This is because of the small couplings of $h_1$, $a_1$, and $\chi_1^0$ to the SM degrees of freedom in this scenario and the small amount of mixing of both $h^0_v$ and $h^0_s$ with $H^0_v$.  Also, all LEP and Tevatron constraints on a light $a_1$ are easily satisfied in this case due to its singlet-like nature.  Furthermore, the down-type couplings of $h_1$ are small in this case, and so bounds from $\Upsilon$ decays are easily satisfied (the corresponding bounds from $\Upsilon\rightarrow a_1 \gamma$ are satisfied since $m_{a_1}>m_{\Upsilon}$). 

The properties of $h_2$ in this scenario are very SM-like, with a substantially reduced invisible branching fraction and total Higgs total width as compared to the light CP-even case.  In particular, both the partial widths for the exotic decays $h_2\rightarrow h_1 h_1$, $h_2\rightarrow \chi_1^0, \chi_1^0$ are small, since they are governed by the couplings $g_{h_2 h_1 h_1}$, which is reduced for $\kappa\ll \lambda\ll1$ (see Eq.~(\ref{eq:h2h1h1})), and $g_{h_2\chi \chi}$, respectively, with the latter given by
\begin{equation}
g_{h_2\chi\chi}\hspace{.1cm}\approx \hspace{.1cm} -\sqrt{2}\kappa N_{15}^2
\end{equation}
in this case, which is minimized for small $\kappa$.  Also, the reduced couplings of $h_2$ to SM degrees of freedom are all very close to unity across the parameter space; in particular, $\kappa_{V}$ falls in the range $0.99\leq \kappa_{V}\leq 1.00$ for all points shown in Fig.~\ref{fig:CPO}.  This scenario features smaller deviations in the $h_2$ $SU(2)$ gauge boson couplings because small values for $\lambda$ and $\kappa$ reduce the size of the off-diagonal entries in $\mathcal{M}_S^2$ (see Eq.~(\ref{eq:MS})).  Consequently, this scenario generally lies well within the current best-fit regions for the 126 GeV Higgs boson in Ref.~\cite{Belanger:2013xza}, in contrast to the light CP-even case which required accidental cancellations in specific regions of the parameter space to achieve agreement of $h_2$ with current observations. 

\begin{table}[tc]
\centering

 \begin{tabular}{|| c || c || c || c || c ||}
\hline \hline
  $\lambda$  &   $\kappa$   & \hspace{.3 cm}  $\tan \beta$ \hspace{.3 cm} &\hspace{.05 cm}  $A_{\lambda}$ [GeV] \hspace{.05 cm} &  \hspace{.05 cm} $A_{\kappa}$ [GeV]  \hspace{.05 cm}\\  \hline

0.26 & 0.0085 & 18 & 3242.0 & -36.0 \\
\hline\hline
$\mu$ [GeV]  \hspace{.05 cm} & \hspace{.05 cm}  $M_1$ [GeV] \hspace{.05 cm}  & \hspace{.05 cm}  $M_2$ [GeV] \hspace{.05 cm}&  \hspace{.05 cm}  $m_{h_1}$ [GeV] \hspace{.05 cm}  & \hspace{.05 cm}  $m_{h_2}$ [GeV] \hspace{.05 cm}    \\
\hline
159 & -131 & 800 & 4.81 & 124.0 \\
\hline \hline
\hspace{.05 cm}  $m_{a_1}$ [GeV] \hspace{.05 cm}  &\hspace{.05 cm}  $m_{a_2}$ [GeV] \hspace{.05 cm}  & \hspace{.05 cm}  $m_{\chi_1}$ [GeV] \hspace{.05 cm}  & $\Omega h^2$ & $\sigma_{\rm SI}$ [cm$^2$]\\
\hline
   27.12 & 2991.7 & 11.5 & 0.132 & $2.2 \times 10^{-42}$  \\
   \hline \hline
$\sigma ^{\rm SD}$ [cm$^2$]&$\sigma v$ [cm$^3/$s] &  $\Gamma_{h_2}^{tot}/\Gamma_{h_{SM}}^{tot}$ & $BR(h_2\rightarrow h_1 h_1)$ & $BR(h_2 \rightarrow \chi_1^0 \chi_1^0)$\\ \hline
 $1.8 \times 10^{-40}$& $1.3 \times 10^{-28}$ & 1.1 & 2.4\% & 3.6\% \\
\hline \hline

\end{tabular}
\caption{\label{table:bm3} \it\small Benchmark point for a light singlino-like neutralino with $\sigma_{\rm SI}$ in the current $2\sigma$ best-fit region for the CDMS II results.  The stop masses are all set to 1 TeV with triscalar couplings $A_{Q,U}=1.8$ TeV.  All couplings of $h_2$, as well as its invisible branching fraction and total width, fall well within the $95\%$ C.L. regions suggested by the global fit in Ref.~\cite{Belanger:2013xza}.  The branching fraction of $h_2$ into light pseudoscalars is also small: $BR(h_2\rightarrow a_1 a_1)=5\%$.}
\end{table}

As a quantitative example, we present  the details for a benchmark point in our scan in Table~\ref{table:bm3}, marked by a black star in Fig.~\ref{fig:CPO}.  This point has a large enough $\sigma_{\rm SI}$ to explain the CDMS II signal, and satisfies the relevant collider, Higgs, and flavor constraints implemented in \texttt{NMSSMTools} and \texttt{HiggsBounds} (implemented in \texttt{MicrOmegas}).  The spin-dependent scattering cross-section is again rather large but experimentally allowed (see e.g. Ref.~\cite{Buckley:2013gjo}).  Despite the large spin-dependent cross-section, LHC Mono-jet search constraints are satisfied by a significant margin, checking against the results of e.g. Ref.~\cite{Cotta:2013jna}.  This is because the decay $h_1\rightarrow \chi_1 \chi_1$ is kinematically forbidden while the product $g_{a_1 \chi \chi} g_{a_1 b \bar{b}}$ is small.  Additionally, the couplings of $h_2$ to the various Standard Model particles, as well as $BR(h_2\rightarrow {\rm invis}.)$ and $\Gamma_{h_2}^{tot}$, fall well within the $95\%$ C.L. regions resulting from the global fit in Ref.~\cite{Belanger:2013xza}.  This point also features a partial width for $h_2\rightarrow \gamma \gamma$, consistent with the SM prediction, with $\kappa_{\gamma}=1.0$.  Note that $\kappa$ can be increased to raise $\sigma_{\rm SI}$ for this benchmark without violating any constraints, although the price paid is a heavier neutralino.

Besides ongoing efforts in dark matter direct detection experiments, the most promising tests of this scenario would seem to be those associated with detecting a light $h_1$, perhaps through $\Upsilon$ decays or other precision measurements.  $B$-physics experiments may play a role, however, as we have seen, these constraints can often be avoided by taking $h_1$ to be heavier than 4.8 GeV.  Collider searches for chargino and neutralino signatures can also be important discriminators in the future, but again the precise signatures expected depend on the details of the spectrum in which there is ample freedom.  Future work is needed to address how effective such searches can be in probing the light CP-even/CP-odd case.  Since the couplings of $h_2$ tend not to deviate substantially from those predicted for the Standard Model Higgs in the light CP-even/CP-odd scenario, this case will be more difficult to confirm or exclude than the light CP-even scenario from LHC Higgs considerations alone, likely requiring precise determinations of the 126 GeV Higgs boson couplings and properties to be able to draw any final conclusions.  Given the current status of such measurements at the LHC by ATLAS and CMS, the light CP-even/CP-odd scenario may be posed to remain a viable dark matter explanation of the CDMS II results for some time to come.

\section{Discussion and Conclusions} \label{sec:conc}

In this work, we have studied the possibility of accommodating and explaining certain direct detection signals with thermal relic neutralino dark matter in the context of the next-to-minimal supersymmetric extension of the Standard Model. We have argued, and demonstrated, that LHC Higgs studies pose significant challenges to this scenario, significantly altering the pre-LHC picture, but also that ongoing and future exploration of the Higgs sector could uniquely unveil the setup under consideration.

With a detailed analysis of the relevant NMSSM parameter space, we have shown that there generically exist at least two physical scenarios that might produce a dark matter candidate with a large enough neutralino-proton scalar scattering cross section to explain the CDMS II events as well as the observed dark matter thermal relic density. In the first, the particle responsible for mediating neutralino pair-annihilation, as well as for producing a large direct detection cross section, is a very light singlet-like CP-even state with a mass in the vicinity of 5 GeV. Besides rare $B$ and $\Upsilon$ decays, LHC data regarding the Higgs sector provides severe constraints in this case, which tends to feature a relatively large Higgs decay branching ratio into pairs of neutralinos and singlet-like lightest Higgs bosons. On the upside, future LHC measurements of the various Higgs production and decay rates will narrow the (inferred) range for the total Higgs decay width and invisible branching fraction and will thereby offer opportunities to directly test this scenario.

In the second physical realization, the thermal relic density is driven down by a relatively light CP-odd Higgs mass eigenstate (on the order of about 30 GeV), while the large neutralino-proton scattering cross section is again enhanced by a very light ($\sim$5 GeV) CP-even singlet-like state. In this case the Higgs branching fractions into lightest neutralino pairs and the two light singlet-like Higgses are also suppressed, resulting in much milder constraints from LHC Higgs results.  $B$- and $\Upsilon$-decay constraints still tightly limit this possibility, however.

Although both scenarios feature a thermal relic light neutralino, which one might naively expect to be in tension with indirect detection constraints, the late-universe pair-annihilation cross section is suppressed in both cases, as the annihilation channels relevant in the early universe at the time of neutralino freeze-out shut down at zero temperature. Postulating non-thermal production mechanisms can, however, change this conclusion for the light CP-even/CP-odd case, although it would also change one of the two key criteria used in selecting the regions of the NMSSM parameter space we considered.

While we were able to find parameter space capable of explaining the CDMS II results, we have generally found it difficult to obtain light enough neutralinos with large enough spin-independent scattering cross-sections to explain the DAMA/LIBRA, CoGeNT, and CRESST-II results ignoring the CDMS II events.  If these ``signals" persist, it may be necessary to look beyond the scale-invariant NMSSM for a compelling explanation.  

The present study highlights the importance of LHC Higgs studies in shedding light on the dark matter sector. While we focused here on a rather model-dependent setup (a perspective that allowed us to draw very specific conclusions about the parameter space and predictions for possible tests of the scenario), we believe one can also extract more general lessons from our findings. In our opinion, the most important such lesson is that precision Higgs studies, especially relating to the invisible and total Higgs decay widths, can be a crucial tool in constraining or honing in on viable dark matter models, especially if the dark matter is light, and if the dark sector is linked to the Standard Model via a light ``Higgs portal''.



\begin{acknowledgments}
\noindent  We thank Patrick Draper and Jason Nielsen for useful discussions.  SP is partly supported by the US Department of Energy under Contract DE-FG02-04ER41268. 
\end{acknowledgments}

\section*{Note Added}

After the first preprint version of our work was posted, Ref.~\cite{Schmidt-Hoberg:2013hba} (and later Ref.~\cite{Clarke:2013aya}) appeared, which highlighted the importance of bounds arising from the exclusive decay $B\rightarrow K \mu^+\mu^-$ mediated by an on-shell light scalar.  These constraints were not included in the first version of our work, although they had also been pointed out previously by e.g. Ref.~\cite{Batell:2009jf}.  In the present version, we have overhauled our analysis, considering heavier singlet-like scalars to evade these constraints. Although limiting ourselves to heavier mediators makes it more difficult to achieve cross-sections in the region suggested by e.g. CoGeNT, our overall conclusions about explaining the CDMS II signal have not changed.  Also, in the meantime, two other experiments, LUX \cite{Lux} and SuperCDMS \cite{Agnese:2014aze}, released even more stringent limits on light WIMPs with large spin-independent scattering cross-sections.  Despite these developments, we believe our study is still important since 1) the picture has not fully settled, with the various `signals' (or anomalies) persisting in the data, and 2) because the work here can be useful in future applications considering a very light scalar in the NMSSM  (or other singlet extensions of the SM) with or without demanding a signal in direct detection experiments.


\begin{thebibliography}{300}

  \bibitem{Hooper:2012ft} 
  D.~Hooper,
  Phys.\ Dark Univ.\  {\bf 1}, 1 (2012)
  [arXiv:1201.1303 [astro-ph.CO]].
  
  \bibitem{Hooper:2011ti} 
  D.~Hooper and T.~Linden,
  Phys.\ Rev.\ D {\bf 84}, 123005 (2011)
  [arXiv:1110.0006 [astro-ph.HE]].
  
  \bibitem{Linden:2011au} 
  T.~Linden, D.~Hooper and F.~Yusef-Zadeh,
  Astrophys.\ J.\  {\bf 741}, 95 (2011)
  [arXiv:1106.5493 [astro-ph.HE]].
  
  \bibitem{Hooper:2012jc} 
  D.~Hooper, A.~V.~Belikov, T.~E.~Jeltema, T.~Linden, S.~Profumo and T.~R.~Slatyer,
  Phys.\ Rev.\ D {\bf 86}, 103003 (2012)
  [arXiv:1203.3547 [astro-ph.CO]].
   
  \bibitem{Bernabei:2010mq} 
  R.~Bernabei {\it et al.}  [DAMA and LIBRA Collaborations],
  Eur.\ Phys.\ J.\ C {\bf 67}, 39 (2010)
  [arXiv:1002.1028 [astro-ph.GA]].
  
  \bibitem{Aalseth:2011wp} 
  C.~E.~Aalseth, P.~S.~Barbeau, J.~Colaresi, J.~I.~Collar, J.~Diaz Leon, J.~E.~Fast, N.~Fields and T.~W.~Hossbach {\it et al.},
  Phys.\ Rev.\ Lett.\  {\bf 107}, 141301 (2011)
  [arXiv:1106.0650 [astro-ph.CO]].
  
  \bibitem{Aalseth:2010vx} 
  C.~E.~Aalseth {\it et al.}  [CoGeNT Collaboration],
  Phys.\ Rev.\ Lett.\  {\bf 106}, 131301 (2011)
  [arXiv:1002.4703 [astro-ph.CO]].
  
  \bibitem{Angloher:2011uu} 
  G.~Angloher, M.~Bauer, I.~Bavykina, A.~Bento, C.~Bucci, C.~Ciemniak, G.~Deuter and F.~von Feilitzsch {\it et al.},
  Eur.\ Phys.\ J.\ C {\bf 72}, 1971 (2012)
  [arXiv:1109.0702 [astro-ph.CO]].
  
 \bibitem{cdms3events}
 R.~Agnese {\it et al.}  [CDMS Collaboration],
  [arXiv:1304.4279 [hep-ex]].
  
  \bibitem{Aprile:2012nq} 
  E.~Aprile {\it et al.}  [XENON100 Collaboration],
  Phys.\ Rev.\ Lett.\  {\bf 109}, 181301 (2012)
  [arXiv:1207.5988 [astro-ph.CO]].
  
  \bibitem{Lux} 
  D.~S.~Akerib {\it et al.}  [LUX Collaboration],
  arXiv:1310.8214 [astro-ph.CO].
  

  \bibitem{Hooper:2013cwa} 
  D.~Hooper,
  arXiv:1306.1790 [hep-ph].
  
  \bibitem{Frandsen:2013cna} 
  M.~T.~Frandsen, F.~Kahlhoefer, C.~McCabe, S.~Sarkar and K.~Schmidt-Hoberg,
  JCAP {\bf 1307}, 023 (2013)
  [arXiv:1304.6066 [hep-ph]].
  
\bibitem{Arbey:2012na} 
  A.~Arbey, M.~Battaglia and F.~Mahmoudi,
  Eur.\ Phys.\ J.\ C {\bf 72}, 2169 (2012)
  [arXiv:1205.2557 [hep-ph]].
  
  \bibitem{Kuflik:2010ah} 
  E.~Kuflik, A.~Pierce and K.~M.~Zurek,
  Phys.\ Rev.\ D {\bf 81}, 111701 (2010)
  [arXiv:1003.0682 [hep-ph]].

  
  \bibitem{Feldman:2010ke} 
  D.~Feldman, Z.~Liu and P.~Nath,
  Phys.\ Rev.\ D {\bf 81}, 117701 (2010)
  [arXiv:1003.0437 [hep-ph]].
  
  \bibitem{Boehm:2013qva} 
  C.~Boehm, P.~S.~B.~Dev, A.~Mazumdar and E.~Pukartas,
  JHEP {\bf 1306}, 113 (2013)
  [arXiv:1303.5386 [hep-ph]].
  
  \bibitem{Calibbi:2013poa} 
  L.~Calibbi, J.~M.~Lindert, T.~Ota and Y.~Takanishi,
  arXiv:1307.4119 [hep-ph].
  
  \bibitem{Arbey:2013aba} 
  A.~Arbey, M.~Battaglia and F.~Mahmoudi,
  arXiv:1308.2153 [hep-ph].
  
  \bibitem{Gondolo:2013wwa}
  P.~Gondolo and S.~Scopel,
  arXiv:1307.4481 [hep-ph].
  
  \bibitem{Bae:2010hr} 
  K.~J.~Bae, H.~D.~Kim and S.~Shin,
  Phys.\ Rev.\ D {\bf 82}, 115014 (2010)
  [arXiv:1005.5131 [hep-ph]].
  
  \bibitem{Gunion:2010dy} 
  J.~F.~Gunion, A.~V.~Belikov and D.~Hooper,
  arXiv:1009.2555 [hep-ph].
  
  \bibitem{DLH} 
  P.~Draper, T.~Liu, C.~E.~M.~Wagner, L.~-T.~Wang and H.~Zhang,
  Phys.\ Rev.\ Lett.\  {\bf 106}, 121805 (2011)
  [arXiv:1009.3963 [hep-ph]].
  
  \bibitem{Belikov:2010yi} 
  A.~V.~Belikov, J.~F.~Gunion, D.~Hooper and T.~M.~P.~Tait,
  Phys.\ Lett.\ B {\bf 705}, 82 (2011)
  [arXiv:1009.0549 [hep-ph]].
  
  \bibitem{Cao:2011re} 
  J.~-J.~Cao, K.~-i.~Hikasa, W.~Wang, J.~M.~Yang, K.~-i.~Hikasa, W.~-Y.~Wang and J.~M.~Yang,
  Phys.\ Lett.\ B {\bf 703}, 292 (2011)
  [arXiv:1104.1754 [hep-ph]].
  
  \bibitem{Cao:2010fi} 
  J.~Cao, K.~-i.~Hikasa, W.~Wang, J.~M.~Yang and L.~-X.~Yu,
  JHEP {\bf 1007}, 044 (2010)
  [arXiv:1005.0761 [hep-ph]].
  
  \bibitem{Vasquez:2010ru} 
  D.~A.~Vasquez, G.~Belanger, C.~Boehm, A.~Pukhov and J.~Silk,
  Phys.\ Rev.\ D {\bf 82}, 115027 (2010)
  [arXiv:1009.4380 [hep-ph]].
  
  \bibitem{AlbornozVasquez:2011js} 
  D.~Albornoz Vasquez, G.~Belanger and C.~Boehm,
  Phys.\ Rev.\ D {\bf 84}, 095008 (2011)
  [arXiv:1107.1614 [hep-ph]].
  
  \bibitem{Vasquez:2012hn} 
  D.~A.~Vasquez, G.~Belanger, C.~Boehm, J.~Da Silva, P.~Richardson and C.~Wymant,
  Phys.\ Rev.\ D {\bf 86}, 035023 (2012)
  [arXiv:1203.3446 [hep-ph]].
  
  \bibitem{godbolepaper}
    G.~Belanger, G.~D.~La Rochelle, B.~Dumont, R.~M.~Godbole, S.~Kraml and S.~Kulkarni,
  arXiv:1308.3735 [hep-ph].
  
    \bibitem{Atlas_couplings} 
  ATLAS-CONF-2013-034 [ATLAS Collaboration]
   
       \bibitem{CMS_couplings} 
 CMS-PAS-HIG-13-005 [CMS Collaboration]
 
  \bibitem{Choi:2013fva} 
  K.~-Y.~Choi and O.~Seto,
  Phys.\ Rev.\ D {\bf 88}, 035005 (2013)
  [arXiv:1305.4322 [hep-ph]].
  
  \bibitem{Cotta:2013jna} 
  R.~C.~Cotta, A.~Rajaraman, T.~M.~P.~Tait and A.~M.~Wijangco,
  arXiv:1305.6609 [hep-ph].
  
  \bibitem{Wang:2012ry} 
  W.~Wang,
  Adv.\ High Energy Phys.\  {\bf 2012}, 216941 (2012)
  [arXiv:1205.5081 [hep-ph]].
  
  \bibitem{Cerdeno:2013cz} 
  D.~G.~Cerdeno, P.~Ghosh and C.~B.~Park,
  arXiv:1301.1325 [hep-ph].
  
  \bibitem{Ellwanger:2009dp} 
  U.~Ellwanger, C.~Hugonie and A.~M.~Teixeira,
  Phys.\ Rept.\  {\bf 496}, 1 (2010)
  [arXiv:0910.1785 [hep-ph]].
  
  \bibitem{Agashe:2012zq} 
  K.~Agashe, Y.~Cui and R.~Franceschini,
  JHEP {\bf 1302}, 031 (2013)
  [arXiv:1209.2115 [hep-ph]].
  
  \bibitem{Feng:2013tvd} 
  J.~L.~Feng, P.~Kant, S.~Profumo and D.~Sanford,
  arXiv:1306.2318 [hep-ph].
 
 \bibitem{Gunion:1984yn} 
  J.~F.~Gunion and H.~E.~Haber,
  Nucl.\ Phys.\ B {\bf 272}, 1 (1986)
  [Erratum-ibid.\ B {\bf 402}, 567 (1993)].
  
  \bibitem{ColemanWeinberg} 
  S.~R.~Coleman and E.~J.~Weinberg,
  Phys.\ Rev.\ D {\bf 7}, 1888 (1973).
  
  \bibitem{Ellis:2000ds} 
  J.~R.~Ellis, A.~Ferstl and K.~A.~Olive,
  Phys.\ Lett.\ B {\bf 481}, 304 (2000)
  [hep-ph/0001005].
  
  \bibitem{PLANCK} 
  P.~A.~R.~Ade {\it et al.}  [PLANCK Collaboration],
  arXiv:1303.5076 [astro-ph.CO].
  
  \bibitem{WMAP9} 
  G.~Hinshaw {\it et al.}  [WMAP Collaboration],
  arXiv:1212.5226 [astro-ph.CO].
  
  \bibitem{Masiero:2004ft} 
  A.~Masiero, S.~Profumo and P.~Ullio,
  Nucl.\ Phys.\ B {\bf 712}, 86 (2005)
  [hep-ph/0412058].
  
  \bibitem{ArkaniHamed:2006mb} 
  N.~Arkani-Hamed, A.~Delgado and G.~F.~Giudice,
  Nucl.\ Phys.\ B {\bf 741}, 108 (2006)
  [hep-ph/0601041].
  
    \bibitem{LEP_chargino} 
  J.~Beringer \textit{et al.} [Particle Data Group Collaboration],
  Phys.\ Rev.\ D {\bf 86}, 010001 (2012).
  
  \bibitem{Barate:2003sz} 
  R.~Barate {\it et al.}  [LEP Working Group for Higgs boson searches and ALEPH and DELPHI and L3 and OPAL Collaborations],
  Phys.\ Lett.\ B {\bf 565}, 61 (2003)
  [hep-ex/0306033].
  
   \bibitem{ALEPH} 
  D.~Buskulic \textit{et al.} (ALEPH Collaboration), 
  Phys. Lett. B {\bf 313} (1993) 312.
  
  \bibitem{Abbiendi:2002qp} 
  G.~Abbiendi {\it et al.}  [OPAL Collaboration],
  Eur.\ Phys.\ J.\ C {\bf 27}, 311 (2003)
  [hep-ex/0206022].
 
  
  \bibitem{Schael:2006cr} 
  S.~Schael {\it et al.}  [ALEPH and DELPHI and L3 and OPAL and LEP Working Group for Higgs Boson Searches Collaborations],
  Eur.\ Phys.\ J.\ C {\bf 47}, 547 (2006)
  [hep-ex/0602042].
  
  \bibitem{Schael:2010aw} 
  S.~Schael {\it et al.}  [ALEPH Collaboration],
  JHEP {\bf 1005}, 049 (2010)
  [arXiv:1003.0705 [hep-ex]].
  
  \bibitem{Abazov:2009yi} 
  V.~M.~Abazov {\it et al.}  [D0 Collaboration],
  Phys.\ Rev.\ Lett.\  {\bf 103}, 061801 (2009)
  [arXiv:0905.3381 [hep-ex]].
  
  \bibitem{HiggsBounds} 
  P.~Bechtle, O.~Brein, S.~Heinemeyer, G.~Weiglein and K.~E.~Williams,
  Comput.\ Phys.\ Commun.\  {\bf 181}, 138 (2010)
  [arXiv:0811.4169 [hep-ph]];
  P.~Bechtle, O.~Brein, S.~Heinemeyer, G.~Weiglein and K.~E.~Williams,
  Comput.\ Phys.\ Commun.\  {\bf 182}, 2605 (2011)
  [arXiv:1102.1898 [hep-ph]];
   P.~Bechtle, O.~Brein, S.~Heinemeyer, O.~Stal, T.~Stefaniak, G.~Weiglein and K.~Williams,
  PoS CHARGED {\bf 2012}, 024 (2012)
  [arXiv:1301.2345 [hep-ph]].
  
    \bibitem{NMSSMTools} 
  U.~Ellwanger, J.~F.~Gunion and C.~Hugonie,
  JHEP {\bf 0502}, 066 (2005)
  [hep-ph/0406215];
  U.~Ellwanger and C.~Hugonie,
  Comput.\ Phys.\ Commun.\  {\bf 175}, 290 (2006)
  [hep-ph/0508022]
  
  \bibitem{Belanger:2013xza} 
  G.~Belanger, B.~Dumont, U.~Ellwanger, J.~F.~Gunion and S.~Kraml,
  arXiv:1306.2941 [hep-ph].
  
  \bibitem{Barger:2012hv} 
  V.~Barger, M.~Ishida and W.~-Y.~Keung,
  Phys.\ Rev.\ Lett.\  {\bf 108}, 261801 (2012)
  [arXiv:1203.3456 [hep-ph]].
  
  \bibitem{Carmi:2012in} 
  D.~Carmi, A.~Falkowski, E.~Kuflik, T.~Volansky and J.~Zupan,
  JHEP {\bf 1210}, 196 (2012)
  [arXiv:1207.1718 [hep-ph]].
  
  \bibitem{Dobrescu:2012td} 
  B.~A.~Dobrescu and J.~D.~Lykken,
  JHEP {\bf 1302}, 073 (2013)
  [arXiv:1210.3342 [hep-ph]].
  
  \bibitem{Bai:2011wz} 
  Y.~Bai, P.~Draper and J.~Shelton,
  JHEP {\bf 1207}, 192 (2012)
  [arXiv:1112.4496 [hep-ph]].
  
  \bibitem{Belanger:2013kya} 
  G.~Belanger, B.~Dumont, U.~Ellwanger, J.~F.~Gunion and S.~Kraml,
  Phys.\ Lett.\ B {\bf 723}, 340 (2013)
  [arXiv:1302.5694 [hep-ph]].
  
  \bibitem{Falkowski:2013dza} 
  A.~Falkowski, F.~Riva and A.~Urbano,
  arXiv:1303.1812 [hep-ph].
  
  \bibitem{LHCb} 
  RAaij {\it et al.}  [LHCb Collaboration],
  JHEP {\bf 1302}, 105 (2013)
  [arXiv:1209.4284 [hep-ex]].
  
  \bibitem{BELLE} 
  J.~-T.~Wei {\it et al.}  [BELLE Collaboration],
  Phys.\ Rev.\ Lett.\  {\bf 103}, 171801 (2009)
  [arXiv:0904.0770 [hep-ex]].
  
  \bibitem{BaBar} 
  J.~P.~Lees {\it et al.}  [BaBar Collaboration],
  Phys.\ Rev.\ D {\bf 86}, 032012 (2012)
  [arXiv:1204.3933 [hep-ex]].
  
  \bibitem{Batell:2009jf} 
  B.~Batell, M.~Pospelov and A.~Ritz,
  Phys.\ Rev.\ D {\bf 83}, 054005 (2011)
  [arXiv:0911.4938 [hep-ph]].
  
  \bibitem{Gunion:1989we} 
  J.~F.~Gunion, H.~E.~Haber, G.~L.~Kane and S.~Dawson,
  Front.\ Phys.\  {\bf 80}, 1 (2000).
  
  \bibitem{ups_hadronic} 
  J.~P.~Lees {\it et al.}  [BaBar Collaboration],
  Phys.\ Rev.\ Lett.\  {\bf 107}, 221803 (2011)
  [arXiv:1108.3549 [hep-ex]].
  
  \bibitem{ups_tau} 
  J.~P.~Lees {\it et al.}  [BaBar Collaboration],
  Phys.\ Rev.\ D {\bf 88}, 071102 (2013)
  [arXiv:1210.5669 [hep-ex], arXiv:1210.5669 [hep-ex]].
  
  \bibitem{ups_muon} 
  J.~P.~Lees {\it et al.}  [BaBar Collaboration],
  Phys.\ Rev.\ D {\bf 87}, no. 3, 031102 (2013)
  [arXiv:1210.0287 [hep-ex]].
  
  \bibitem{ups_gluon} 
  J.~P.~Lees {\it et al.}  [BaBar Collaboration],
  Phys.\ Rev.\ D {\bf 88}, 031701 (2013)
  [arXiv:1307.5306 [hep-ex]].
  
  \bibitem{McKeen:2008gd} 
  D.~McKeen,
  Phys.\ Rev.\ D {\bf 79}, 015007 (2009)
  [arXiv:0809.4787 [hep-ph]].
  
  \bibitem{Aubert:2009cp} 
  B.~Aubert {\it et al.}  [BaBar Collaboration],
  Phys.\ Rev.\ Lett.\  {\bf 103}, 081803 (2009)
  [arXiv:0905.4539 [hep-ex]].
  
  \bibitem{ATLAS:2012ky} 
  G.~Aad {\it et al.}  [ATLAS Collaboration],
  JHEP {\bf 1304}, 075 (2013)
  [arXiv:1210.4491 [hep-ex]].
  
  \bibitem{Chatrchyan:2012me} 
  S.~Chatrchyan {\it et al.}  [CMS Collaboration],
  JHEP {\bf 1209}, 094 (2012)
  [arXiv:1206.5663 [hep-ex]].
  
  \bibitem{ALEPH:2005ab} 
  S.~Schael {\it et al.}  [ALEPH and DELPHI and L3 and OPAL and SLD and LEP Electroweak Working Group and SLD Electroweak Group and SLD Heavy Flavour Group Collaborations],
  Phys.\ Rept.\  {\bf 427}, 257 (2006)
  [hep-ex/0509008].
  
  \bibitem{Madgraph} 
  J.~Alwall, M.~Herquet, F.~Maltoni, O.~Mattelaer and T.~Stelzer,
  JHEP {\bf 1106}, 128 (2011)
  [arXiv:1106.0522 [hep-ph]].

\bibitem{ATLAS_EW}
  ATLAS Collaboration, ATLAS-CONF-2013-035
  
\bibitem{CMS_EW}
  CMS Collaboration, CMS-PAS-SUS-13-006

  \bibitem{Kanehata:2010ci} 
  Y.~Kanehata, T.~Kobayashi, Y.~Konishi and T.~Shimomura,
  Phys.\ Rev.\ D {\bf 82}, 075018 (2010)
  [arXiv:1008.0593 [hep-ph]].
  
  \bibitem{Kobayashi:2012xv} 
  T.~Kobayashi, T.~Shimomura and T.~Takahashi,
  Phys.\ Rev.\ D {\bf 86}, 015029 (2012)
  [arXiv:1203.4328 [hep-ph]].

\bibitem{Drees:1992am} 
  M.~Drees and M.~M.~Nojiri,
  Phys.\ Rev.\ D {\bf 47}, 376 (1993)
  [hep-ph/9207234].
  
  \bibitem{Gondolo:1990dk} 
  P.~Gondolo and G.~Gelmini,
  Nucl.\ Phys.\ B {\bf 360}, 145 (1991).
 
  \bibitem{light_staus} 
  M.~Carena, S.~Gori, N.~R.~Shah, C.~E.~M.~Wagner and L.~-T.~Wang,
  JHEP {\bf 1207}, 175 (2012)
  [arXiv:1205.5842 [hep-ph]].
  
  \bibitem{Buckley:2013gjo} 
  M.~R.~Buckley and W.~H.~Lippincott,
  Phys.\ Rev.\ D {\bf 88}, 056003 (2013)
  [arXiv:1306.2349 [hep-ph]].
  
    \bibitem{slides} 
   \url{https://indico.fnal.gov/getFile.py/access?contribId=54&sessionId=1&resId=0&materialId=slides&confId=6969}

    \bibitem{Ackermann:2012qk} 
  M.~Ackermann {\it et al.}  [LAT Collaboration],
  Phys.\ Rev.\ D {\bf 86}, 022002 (2012)
  [arXiv:1205.2739 [astro-ph.HE]].
  
  \bibitem{Ando:2012vu} 
  S.~Ando and D.~Nagai,
  JCAP {\bf 1207}, 017 (2012)
  [arXiv:1201.0753 [astro-ph.HE]].
  
  \bibitem{Arcadi:2011ev} 
  G.~Arcadi and P.~Ullio,
  Phys.\ Rev.\ D {\bf 84}, 043520 (2011)
  [arXiv:1104.3591 [hep-ph]].

\bibitem{Schmidt-Hoberg:2013hba} 
  K.~Schmidt-Hoberg, F.~Staub and M.~W.~Winkler,
  Phys.\ Lett.\ B {\bf 727}, 506 (2013)
  [arXiv:1310.6752 [hep-ph]].
  
  \bibitem{Clarke:2013aya} 
  J.~D.~Clarke, R.~Foot and R.~R.~Volkas,
  JHEP {\bf 1402}, 123 (2014)
  [arXiv:1310.8042 [hep-ph]].
  
  \bibitem{Agnese:2014aze} 
  R.~Agnese {\it et al.}  [SuperCDMS Collaboration],
  arXiv:1402.7137 [hep-ex].
  
   \end{thebibliography}
\end{document}